\documentclass[journal, pdftex, draftclsnofoot, onecolumn, 12pt, twoside]{IEEEtran}
\IEEEoverridecommandlockouts
\usepackage{cite}
\usepackage{amsmath,amssymb,amsfonts}
\usepackage{algorithmic}
\usepackage{graphicx}
\usepackage{textcomp}
\usepackage{xcolor}
\usepackage{setspace}
\usepackage{graphicx}
\usepackage{amsthm}
\usepackage{lscape}
\usepackage{latexsym}
\usepackage{bm}
\usepackage{bbm}
\usepackage{cite}
\usepackage{url}
\usepackage{color}
\usepackage[caption=false,font=footnotesize]{subfig}
\usepackage{graphicx}
\usepackage{lmodern}
\usepackage[euler]{textgreek}

\usepackage{pgfplots}
\usepackage{pgfplotstable}
\pgfplotsset{compat=1.17}
\newcounter{marknumber}
\pgfplotsset{
    error bars/every nth mark/.style={
        /pgfplots/error bars/draw error bar/.prefix code={
            \pgfmathtruncatemacro\marknumbercheck{mod(floor(\themarknumber/2),#1)}
            \ifnum\marknumbercheck=0
            \else
                \begin{scope}[opacity=0]
            \fi
        },
        /pgfplots/error bars/draw error bar/.append code={
            \ifnum\marknumbercheck=0
            \else
                \end{scope}
            \fi
            \stepcounter{marknumber}    
        }
    }
}
\def\BibTeX{{\rm B\kern-.05em{\sc i\kern-.025em b}\kern-.08em
    T\kern-.1667em\lower.7ex\hbox{E}\kern-.125emX}}
\begin{document}

\title{DeepJSCC-Q: Constellation Constrained Deep Joint Source-Channel Coding}
\author{Tze-Yang Tung, David Burth Kurka, Mikolaj Jankowski, Deniz Gündüz\\
Department of Electrical and Electronics Engineering\\
Imperial College London
\thanks{This work was supported by the European Research Council (ERC) through project BEACON (No. 677854).}}

\maketitle

\begin{abstract}
Recent works have shown that modern machine learning techniques can provide an alternative approach to the long-standing joint source-channel coding (JSCC) problem. 
Very promising initial results, superior to popular digital schemes that utilize separate source and channel codes, have been demonstrated for wireless image and video transmission using deep neural networks (DNNs).
However, end-to-end training of such schemes requires a differentiable channel input representation; 
hence, prior works have assumed that any complex value can be transmitted over the channel. 
This can prevent the application of these codes in scenarios where the hardware or protocol can only admit certain sets of channel inputs, prescribed by a digital constellation.
Herein, we propose \emph{DeepJSCC-Q}, an end-to-end optimized JSCC solution for wireless image transmission using a finite channel input alphabet. 
We show that \emph{DeepJSCC-Q} can achieve similar performance to prior works that allow any complex valued channel input, especially when high modulation orders are available, and that the performance asymptotically approaches that of unconstrained channel input as the modulation order increases.
Importantly, \emph{DeepJSCC-Q} preserves the graceful degradation of image quality in unpredictable channel conditions, a desirable property for deployment in mobile systems with rapidly changing channel conditions.
\end{abstract}


\section{Introduction}
\label{sec:intro}

\begin{figure}
\begin{center}
\includegraphics[width=0.7\textwidth]{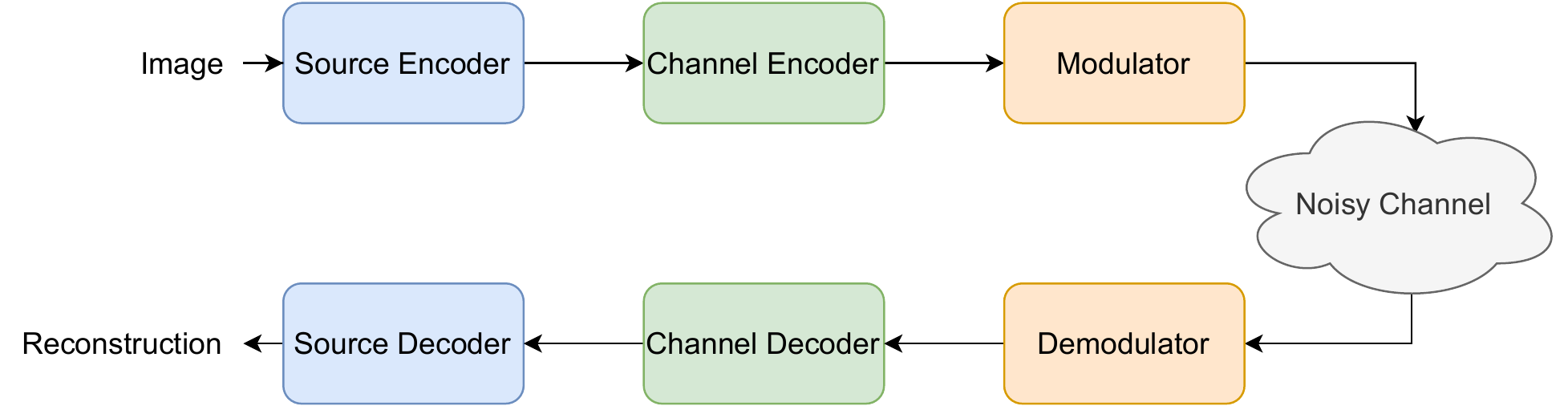}
\end{center}
\caption{Diagram of a typical separation-based digital image communication system.}
\label{fig:separation}
\end{figure}

Source coding and channel coding are two essential steps in modern data transmission. 
The former reduces the redundancy within the source signal, preserving essential information needed to reconstruct the signal within a certain fidelity. 
For example, in image transmission, commonly used compression schemes, such as JPEG and BPG, allow for the reduction in communication load with minimal loss in reconstruction quality.
Channel coding, on the other hand, introduces structured redundancy to allow reliable decoding in the presence of channel imperfections.
A diagram of a typical communication system employing separate source and channel coding is shown in Fig. \ref{fig:separation}.

It was proven by Shannon that the separation of source and channel coding is without loss of optimality when the blocklength goes to infinity \cite{Shannon:1948}. 
Nevertheless, in practical applications we are limited to finite blocklengths, and it is known that combining the two coding steps, that is, joint source-channel coding (JSCC), can achieve lower distortion for a given finite blocklength than separate source and channel coding \cite{gallager_information_1968, kostina_lossy_2013, gastparCodeNotCode2003a}. 
The most straightforward approach to JSCC is to optimize the various parameters between source coding, channel coding and modulation in a cross-layer framework.
Although many such schemes have been proposed over the years \cite{yuEnergyEfficientJPEG2004,appadwedulaJointSourceChannel1998,caiRobustJointSourcechannel2000}, none were able to provide sufficient gains to justify the increased complexity.

A more fundamental approach is to redesign a JSCC scheme from scratch, directly mapping the source signal to the modulated channel input, without conversion to bits at all.
It was shown recently in \cite{Eirina:TCCN:19}, that deep neural networks (DNNs) can be used to break the complexity barrier of designing JSCC schemes for wireless image transmission.
The scheme, called \emph{DeepJSCC}, showed appealing properties, such as lower end-to-end distortion for a given channel blocklength compared with state-of-the-art digital compression schemes \cite{Kurka:IZS2020},
flexibility to adapt to different source or channel models \cite{Eirina:TCCN:19,Kurka:IZS2020}, 
ability to exploit channel feedback \cite{Kurka:deepjsccf:jsait}, 
and capability to produce adaptive-bandwidth transmission schemes \cite{Kurka:BandwidthAgile:TWComm2021}.
Importantly, graceful degradation of image quality with respect to decreasing channel quality means that DeepJSCC is able to avoid the \textit{cliff-effect} that all separation-based schemes suffer from;
which refers to the phenomenon where the image becomes un-decodable when the channel quality falls below a certain threshold resulting in unreliable transmission.

Another strength of DeepJSCC is that it learns a communication scheme from scratch, optimizing all transformations in a data-driven manner using autoencoders \cite{Ballard:Autoencoders:AAAI1987} with a non-trainable differentiable channel model in the bottleneck layer. 
This simplifies the JSCC design procedure, and allows adaptation to any particular source or channel domain and quality measure.
Part of that simplification stems from the fact that DeepJSCC not only combines source and channel coding into one single mapping, but it also removes the constellation diagrams used in digital schemes. 
In digital communications, channel encoded bits are mapped to the elements of a two-dimensional finite constellation diagram, such as quadrature amplitude modulation (QAM), phase shift keying (PSK), or amplitude shift keying (ASK).
In contrast, in DeepJSCC, the encoder can transmit arbitrary complex-valued channel symbols, within a power constraint.
However, this can hinder the adoption of DeepJSCC in current commercial hardware and standardized protocols, which are constrained to produce fixed sets of symbols.

In this work, we investigate the effects of constraining the transmission either to a limited number of channel input symbols, or to a predefined constellation imposed externally. 
This constraint can be crucial for the adoption of DeepJSCC in commercially available hardware (e.g., radio transmitters), where modulators are hard-coded for efficiency, and limit the output space available to the encoder. 
Successfully incorporating fixed channel input constellations in DNN driven JSCC may even open the possibility of incorporating such schemes into established standards, such as 5G telecommunications.
Therefore, in this paper we introduce a new strategy for JSCC of images, called \emph{DeepJSCC-Q}, which allows for the transmission of the content through fixed pre-defined constellations. 

We note that the problem at hand is a JSCC problem over a discrete-input additive white Gaussian noise (AWGN) channel.
For any given finite blocklength, we have a finite number of codewords that can be transmitted.
Hence, the goal is to find the mapping from the input images to these codewords and a matching decoder mapping that minimizes the average end-to-end distortion.
However, finding these mappings directly is a formidable challenge.
We formulate this problem as an end-to-end JSCC problem with a quantizer in the middle.
An encoder DNN extracts the features of the input image, which are then quantized to the constellation points.
These constellation points are transmitted over the channel and the receiver tries to recover the input image from its noisy observations using another DNN.
Hence, the problem becomes the training of an autoencoder architecture with a non-differentiable quantization layer followed by a non-trainable channel layer in the middle.

The main contributions of \emph{DeepJSCC-Q} are:
\begin{itemize}
    \item Achieve performance close to that is achieved by the unconstrained DeepJSCC scheme \cite{Eirina:TCCN:19} even when using a highly constrained channel input representation.
    \item Achieve superior performance compared to separate source and channel coding using better portable graphics (BPG) codec \cite{Bellard:BPG} followed by low density parity check (LDPC) codes \cite{gallager_low-density_1962}.
    \item Create a coherent mapping between the input image and constellation points, avoiding the \textit{cliff-effect} present in all separation-based schemes.
    \item Generate new constellations for a given modulation order, outperforming conventional constellation designs.
\end{itemize}

\section{Related Works}
\label{sec:related_works}

JSCC of images for wireless transmission has received numerous attention over the years.
Earliest communication systems were purely analog, where the source signal is directly modulated unto the carrier waveform, corresponding to a most direct form of JSCC.
With the advances in digital communication and compression techniques, separation based communication systems became dominant.
In order to overcome the limitations of the separation approach, many works focus on optimizing the various parameters of the employed source and channel codes with the goal of minimizing the end-to-end distortion.
For example, in \cite{appadwedulaJointSourceChannel1998}, a general framework for matching source and channel code rates using a parametric distortion model was proposed.
Their approach is to match the source code rate to the channel code and channel statistics in a source-rate-based optimization approach.
Similarly, in \cite{yuEnergyEfficientJPEG2004}, a cross-layer optimization of the source code rate, channel code rate and transmitter power for quality of service (QoS) is proposed.

A slightly different approach considers unequal error protection (UEP) to achieve reliability despite channel uncertainty.
This approach typically separates the source into a base layer and potentially multiple enhancement layers, with the base layer given the greatest amount of error protection to ensure a baseline reconstruction quality. 
In \cite{caiRobustJointSourcechannel2000}, the source image is split into multiple layers in the discrete wavelet transform (DWT) domain before a quantizer and channel code is applied.
The bit rate allocation between the quantizer and channel code of each layer is optimized using an end-to-end rate distortion model.
Similarly, in \cite{thomosWirelessImageTransmission2005}, turbo codes \cite{berrouOptimumErrorCorrecting1996} and Reed-Solomon codes \cite{lin2004error} are used to achieve UEP.
In \cite{arslanCodedHierarchicalModulation2011}, instead of using different channel codes, the authors propose hierarchical modulation to achieve UEP.
However, none of these schemes were able to achieve adequate gains to justify the increase in system complexity stemming from the joint optimization of the parameters of multiple codes and the successive decoding needed at the receiver.

% A more fundamental approach, using low complexity methods to map image pixels directly to channel input symbols was first introduced by \cite{SoftCast:Allerton:10}.
% They exploit the frequency domain sparsity of the images to achieve bandwidth reduction and showed that such schemes can achieve graceful degradation of image quality with respect to channel quality.
% Since then, 
% In \cite{Tung:CL:18}, a low complexity scheme that maps image pixels directly to channel input symbols was introduced.
% It exploits the frequency domain sparsity of the images to achieve bandwidth reduction.
% Although it was shown that this scheme can achieve graceful degradation of image quality with respect to channel quality, it is not competitive to separation based schemes in terms of image quality and cannot and cannot exploit the available bandwidth, or adapt to channel and network conditions dynamically.

A more fundamental approach is to design the communication system from scratch, without considering any digital interface.
Different from the pure analog modulation schemes, source signal is sampled and transmitted using modulated pulses.
However, mapping the source samples to channel inputs is in general a difficult multi-dimensional optimization problem.
Recently, it was shown that DNNs can be used to break the complexity barrier of designing JSCC for wireless image transmission \cite{Eirina:TCCN:19}.
By setting up the encoder and decoder in an autoencoder configuration with a non-trainable channel layer in between, and by using the mean-squared error (MSE) metric as the loss function between the input and the output, the DNN encoder was able to learn a function which maps input images to channel inputs, and vice versa at the decoder. 
They demonstrated that the resultant JSCC encoder and decoder, called \emph{DeepJSCC}, was able to surpass the performance of JPEG2000 \cite{christopoulos_jpeg2000_2000} compression followed by LDPC codes \cite{gallager_low-density_1962} for channel coding.
Importantly, they also showed that such schemes can avoid the \textit{cliff-effect} exhibited by all separation-based schemes, which is when the channel quality deteriorates below the minimum channel quality to allow successful decoding of the deployed channel code, leading to a cliff-edge drop-off in the end-to-end performance.
Since then, various works have extended this result to further demonstrate the ability to exploit channel feedback \cite{Kurka:deepjsccf:jsait} and adapt to various bandwidth requirements without retraining \cite{Kurka:BandwidthAgile:TWComm2021}.
In \cite{yang_deep_2021}, the viability of DeepJSCC in an orthogonal frequency division multiplexing (OFDM) system was shown, and in \cite{ding_snr-adaptive_2021} it was shown that such schemes can be extended to multi-user scenarios using the same decoder.
DNN-aided JSCC for wireless video transmission is studied in \cite{tungDeepWiVeDeepLearningAidedWireless2021}.

However, an implicit assumption in all of the works above is the ability for the communication hardware to transmit arbitrary complex valued channel inputs.
This may not be true as many commercially available hardware have hard-coded standard protocols, making these methods less viable for real world deployment.
As such, in \cite{choi_necst_2018}, an image transmission problem is considered over a discrete input channel, where the input representation is learned using a variational autoencoder (VAE) by assuming a Bernoulli prior.
They consider the transmission of MNIST images \cite{deng_mnist_2012} over a binary erasure channel (BEC) and showed that their scheme performs better than a VAE that only performs compression paired with an LDPC code.
An extension to this work \cite{songInfomaxNeuralJoint2020} improved the results by using an adversarial loss function.
In contrast to these works, we investigate the transmission of natural images over a differentiable channel model with a finite channel input alphabet. 

The optimization of the input constellation of a digital communication system has also been a long standing research challenge in information and communication theory \cite{kayhan_constellation_2012,kayhan_constellation_2013,foschini_optimization_1974,foschini_selection_1973,kearsley_global_2001}.
From a channel coding perspective, the goal is to maximize the mutual information between the channel input and output over the distribution of the channel input alphabet. 
This is inline with Shannon's theorem, which states that the capacity of the channel is defined over the channel input distribution.
Although these methods tend to use hand-crafted designs in the past, more recently, DNNs have also been used to carry out constellation learning. 
For example, in \cite{stark_joint_2019}, the authors consider the transmission of a uniform binary source using only a fixed set of channel input alphabet, and show that the probability distribution of the constellation points can be learned by optimizing the bit error rate for a given channel signal-to-noise ratio (SNR).
They also showed that the constellation points themselves can be part of the trainable parameters, resulting in even better performance.
Herein, we are concerned with optimizing the channel input constellation and distribution for natural image sources, rather than the transmission of uniform binary data sequences.

\section{Problem Statement}
\label{sec:problem_def}

\begin{figure}
\centering
\subfloat[Scenario 1: both the transmitter and receiver have full CSI knowledge.]{
\includegraphics[width=0.46\textwidth]{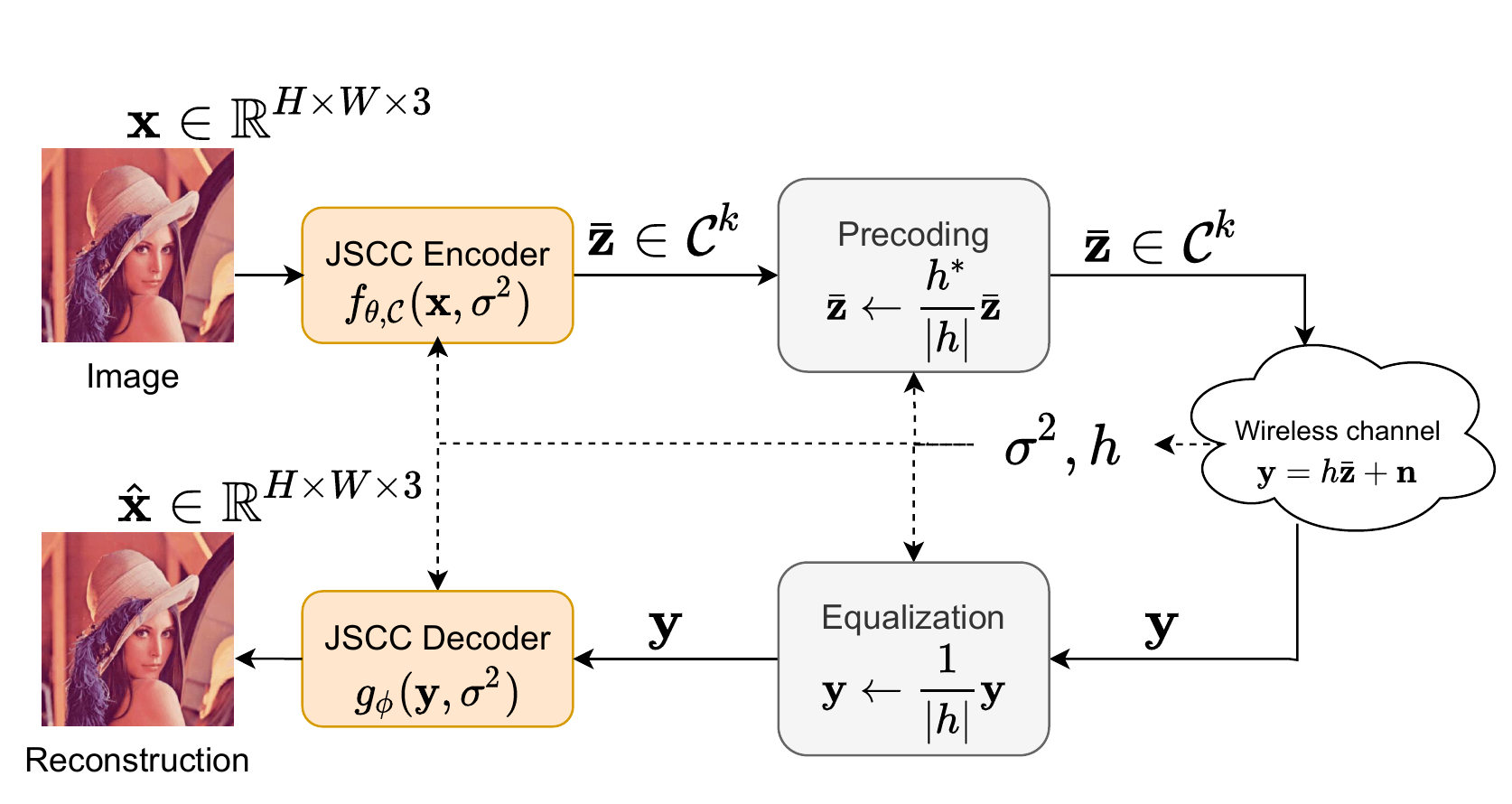}
}
\hfill
\subfloat[Scenario 2: the transmitter knows the noise power while the receiver has full CSI knowledge.]{
\includegraphics[width=0.46\textwidth]{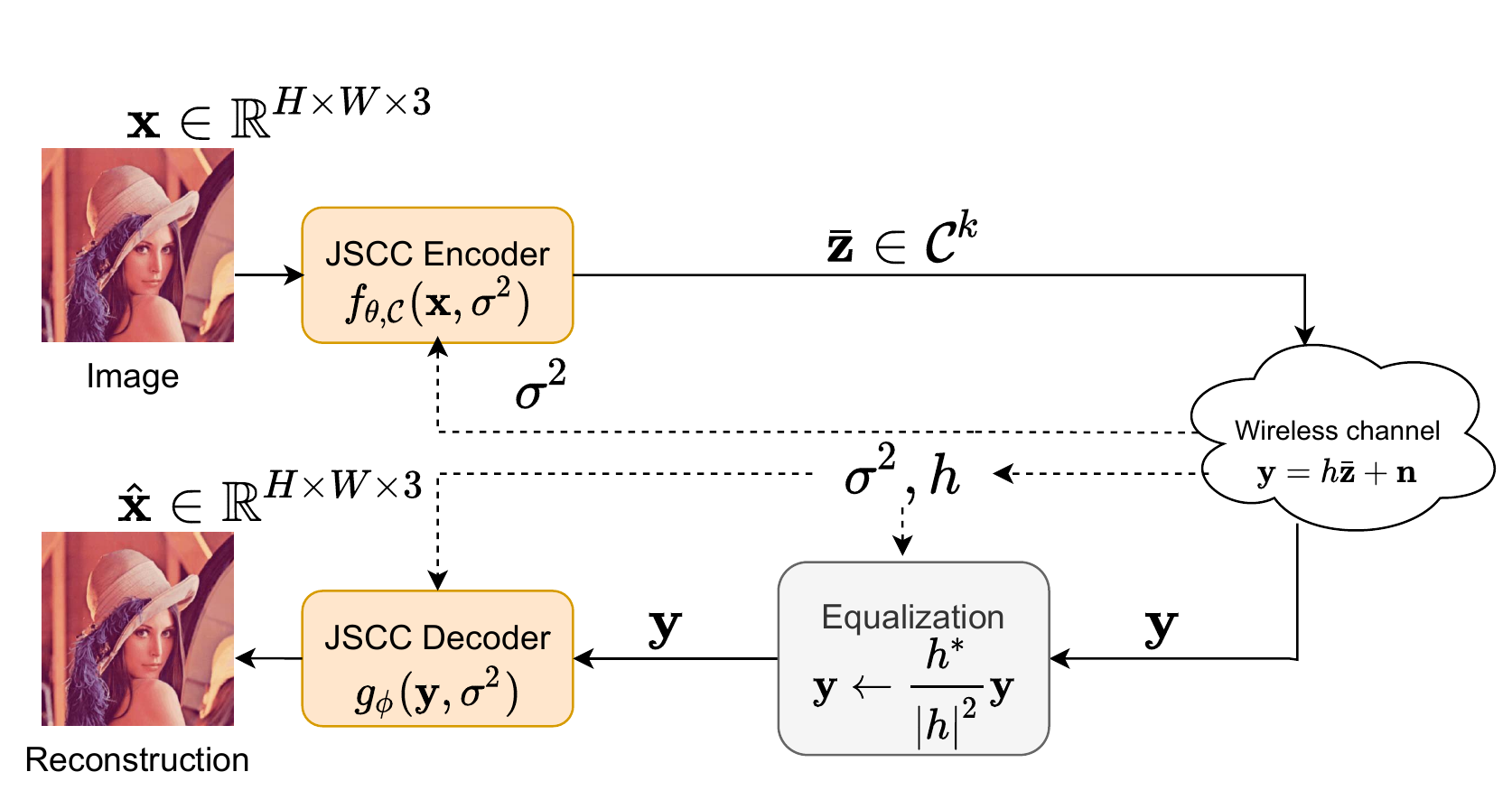}
}
\caption{Overview of problem definition for both CSI scenarios.}
\label{fig:problem_def}
\end{figure}

We consider the problem of wireless image transmission over a noisy channel, in which communication is performed by transmitting one out of a finite set of symbols at each channel use.
An input image $\mathbf{x} \in \{0,...,255\}^{H\times W\times C}$ (where $H$, $W$ and $C$ represent the image's height, width and color channels, respectively) is mapped with an encoder function $f:\{0,...,255\}^{H\times W\times C} \mapsto \mathcal{C}^k$, where $\mathcal{C} = \{c_1,...,c_M\} \subset \mathbb{C}$ is the channel input alphabet with $|\mathcal{C}|=M$.
We impose an average transmit power constraint $\bar{P}$, such that
\begin{equation}
    \frac{1}{k}\sum_{i=1}^k |\bar{z}_i|^2 \leq \bar{P},
\end{equation}
where $\bar{z}_i$ is the $i$th element of vector $\bar{\mathbf{z}}$.
The channel input vector $\bar{\mathbf{z}}$ is transmitted through a noisy channel, with the transfer function $\mathbf{y} = \Upsilon(\bar{\mathbf{z}}) =  h \bar{\mathbf{z}} + \mathbf{n}$, where $h \in \mathbb{C}$ is the channel gain and $\mathbf{n} \sim  CN(0,\sigma^2\mathbf{I}_{k\times k})$ is a complex Gaussian vector with dimensionality $k$. 
% Without loss of generality, we will assume the channel gain $\mathbf{H}$ is diagonal $\text{diag}(\mathbf{H}) = (h, ..., h)$, where $h \sim CN(0, 1)$ for $i = 1, ..., k$.
Finally, a receiver passes the channel output through a decoder function $g:\mathbb{C}^k \mapsto \{0,...,255\}^{H\times W\times C}$ to produce a reconstruction of the input $\hat{\mathbf{x}}=g(\mathbf{y})$.

We consider both the static and fading channel scenarios.
In the former, $h$ is constant, and it is known by both the transmitter and the receiver, corresponding to an AWGN channel.
In the case of a fading channel, we assume that $h$ takes on an independent value during the transmission of each image, and its realization is known only by the receiver.
Since the channel state information (CSI) is known by the receiver, it can perform channel equalization as
\begin{equation}
    \mathbf{y} \leftarrow \frac{h^\ast}{|h|^2} \mathbf{y},
    \label{eq:equalization}
\end{equation}
where $h^\ast$ is the complex conjugate of $h$.
The equalized symbols are then passed to the decoder for decoding.
If the transmitter also has knowledge of $h$, as in the case of a static channel, then the transmitter can perform precoding, such that 
\begin{equation}
    \bar{\mathbf{z}} \leftarrow \frac{h^\ast}{|h|} \bar{\mathbf{z}}.
\end{equation}
After channel equalization at the receiver, we equivalently obtain an AWGN channel with signal-to-noise ratio (SNR) 
\begin{equation}
    \text{SNR} = 10\log_{10}\left(\frac{|h|^2 \bar{P}}{{\sigma}^2}\right) dB.
\end{equation}
For the fading scenario, the average channel SNR is given by
\begin{equation}
    \text{SNR} = 10\log_{10}\left(\frac{\mathbb{E}[|h|^2] \bar{P}}{{\sigma}^2}\right) dB.
\end{equation}
A diagram illustrating both scenarios is shown in Fig. \ref{fig:problem_def}.
The \emph{bandwidth compression ratio} is defined as
\begin{equation}
    \rho = \frac{k}{H\times W\times C}~\text{channel symbols/pixel},
\end{equation}
which measures how much compression we apply to the images, with smaller number reflecting more compression.

To measure the reconstruction quality, we use two metrics: peak signal-to-noise ratio (PSNR) and multi-scale structural similarity index (MS-SSIM).
They are defined as
\begin{equation}
    \text{PSNR}(\mathbf{x},\hat{\mathbf{x}})= 10\log_{10}\bigg(\frac{A^2}{\text{MSE}(\mathbf{x},\hat{\mathbf{x}})}\bigg)~\text{dB},
    \label{eq:psnr_def}
\end{equation}
where $\text{MSE}(\mathbf{x},\hat{\mathbf{x}})=||\mathbf{x}-\hat{\mathbf{x}}||_2^2$ and $A$ is the maximum possible value for a given pixel. 
For a 24-bit RGB pixel, $A=255$.

The multi-scale structural similarity index (MS-SSIM) is defined as:
\begin{equation}
\begin{aligned}
    \text{MS-SSIM}&(\mathbf{x},\hat{\mathbf{x}}) =\\
    & [l_M(\mathbf{x},\hat{\mathbf{x}})]^{\alpha_M}\prod_{j=1}^M[a_j(\mathbf{x},\hat{\mathbf{x}})]^{\beta_j}[b_j(\mathbf{x},\hat{\mathbf{x}})]^{\gamma_j},
\end{aligned}
    \label{eq:msssim_def}
\end{equation}
where
\begin{align}
    l_M(\mathbf{x},\hat{\mathbf{x}}) = \frac{2\mu_{\mathbf{x}}\mu_{\hat{\mathbf{x}}}+v_1}{\mu_{\mathbf{x}}^2+\mu_{\hat{\mathbf{x}}}^2+v_1},\\
    a_j(\mathbf{x},\hat{\mathbf{x}}) = \frac{2\sigma_{\mathbf{x}}\sigma_{\hat{\mathbf{x}}}+v_2}{\sigma_{\mathbf{x}}^2+\sigma_{\hat{\mathbf{x}}}^2+v_2},\\
    b_j(\mathbf{x},\hat{\mathbf{x}}) = \frac{\sigma_{\mathbf{x}\hat{\mathbf{x}}}+v_3}{\sigma_{\mathbf{x}}\sigma_{\hat{\mathbf{x}}}+v_3},
\end{align}
$\mu_{\mathbf{x}}$, $\sigma^2_{\mathbf{x}}$, $\sigma^2_{\mathbf{x}\hat{\mathbf{x}}}$ are the mean and variance of $\mathbf{x}$, and the covariance between $\mathbf{x}$ and $\hat{\mathbf{x}}$, respectively.
$v_1$, $v_2$, and $v_3$ are coefficients for numeric stability; 
$\alpha_M$, $\beta_j$, and $\gamma_j$ are the weights for each of the components.
Each $a_j(\cdot,\cdot)$ and $b_j(\cdot,\cdot)$ is computed at a different downsampled scale of $\mathbf{x}$ and $\hat{\mathbf{x}}$.
We use the default parameter values of ($\alpha_M$, $\beta_j$, $\gamma_j$) provided by the original paper \cite{wang_multiscale_2003}.
MS-SSIM has been shown to perform better in approximating the human visual perception than the more simplistic structural similarity index (SSIM) on different subjective image and video databases.

The overall goal of our design is to characterize the encoding and decoding functions $f$ and $g$ that maximize the average reconstructed image quality, measured by either Eqn. (\ref{eq:psnr_def}) or (\ref{eq:msssim_def}), between the input image $\mathbf{x}$ and its reconstruction at the decoder $\hat{\mathbf{x}}$, under the given constraints on the available bandwidth ratio $\rho$, average power $P$, and constellation $\mathcal{C}$.

\section{Proposed Solution}
\label{sec:proposed_solution}

\begin{figure}
\begin{center}
\includegraphics[width=\textwidth]{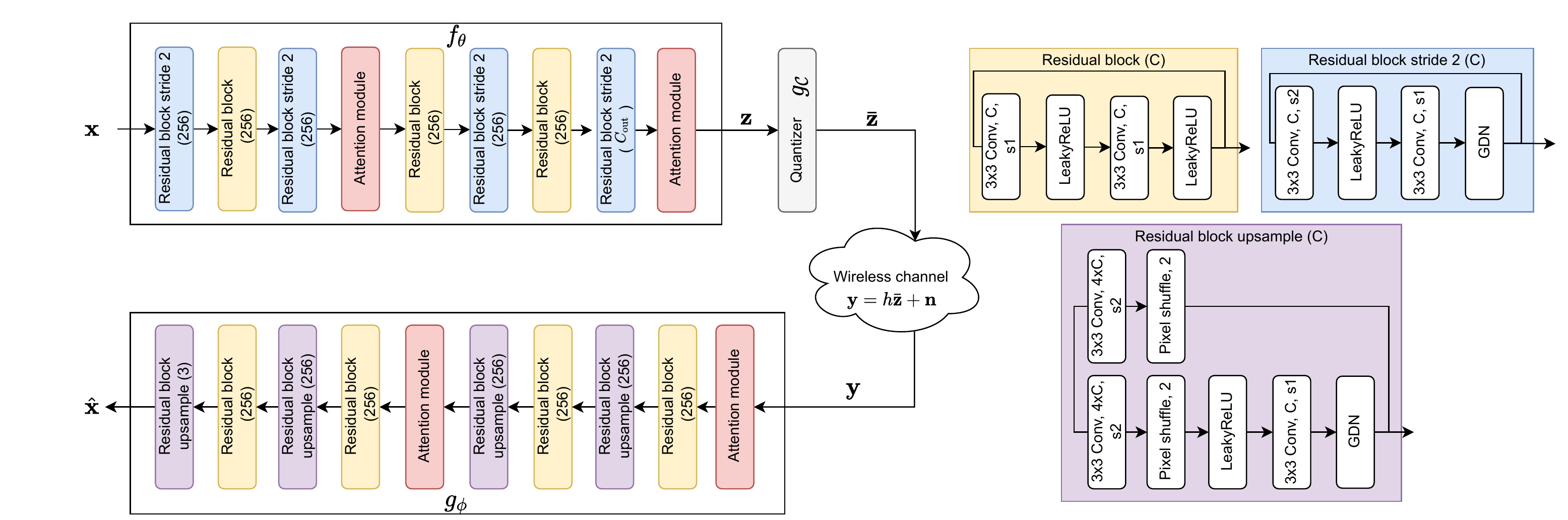}
\end{center}
  \caption{Architecture of the proposed encoder and decoder models.}
\label{fig:architecture}
\end{figure}

Herein, we propose \emph{DeepJSCC-Q}, a DNN-based JSCC scheme and an end-to-end training strategy. 
In \emph{DeepJSCC-Q} we model the encoder and decoder functions as DNNs parameterized by $\boldsymbol{\theta}$ and $\boldsymbol{\phi}$, respectively, and aim at learning the optimal parameters through training. 
Rather than constraining the encoder DNN to discrete outputs, which would require a huge output space, we will allow any output vector of dimension $k$, and employ a ``quantization" layer to map the generated latent vectors to transmitted symbols, such that each quantization level represents a point in the constellation. 
We will introduce two quantization strategies, one where the constellation $\mathcal{C}$ is fixed, and another, where the constellation is also part of the parameters to be trained for a given constellation order $M$.
By making the constellation part of the trainable parameters, we can also optimize the channel input geometry.
As such, we separate the encoder $f$ into two stages: 
first a DNN function $f_{\boldsymbol{\theta}}: \{0,...,255\}^{H \times W \times C} \mapsto \mathbb{C}^k$ maps an input image $\mathbf{x}$ to a complex latent representation $\mathbf{z} = f_{\boldsymbol{\theta}} (\mathbf{x})$ 
before a quantizer $q_{\mathcal{C}}: \mathbb{C}^k \mapsto \mathcal{C}^k$ maps the latent vector $\mathbf{z}$ to the channel input $\bar{\mathbf{z}} = q_{\mathcal{C}}(\mathbf{z})$.

As in previous works \cite{Kurka:IZS2020,Eirina:TCCN:19}, we utilize an autoencoder architecture to jointly train the encoder and the decoder.
We propose a fully convolutional encoder $f_{\boldsymbol{\theta}}$ and decoder $g_{\boldsymbol{\phi}}$ architecture as shown in Fig. \ref{fig:architecture}. 
In the architecture, $C$ refers to the number of channels in the output tensor of the convolution operation.
$C_{\text{out}}$ refers to the number of channels in the final output tensor of the encoder $f_{\boldsymbol{\theta}}$, which controls the number of channel uses $k$ per image.
The ``Pixel shuffle" module, within the ``Residual block upsample" module, is used to increases the height and width of the input tensor by reshaping the it, such that the channel dimensions are reduced while the height and width dimensions are increased. 
This was first proposed in \cite{shi_real-time_2016} as a less computationally expensive method for increasing the CNN tensor dimensions without requiring large number of parameters, like tranpose convolutional layers.
The GDN layer refers to \emph{generalised divisive normalization}, initially proposed in \cite{balle2015density}, and has been shown to be effective in density modeling and compression of images.
The Attention layer refers to the simplified attention module proposed in \cite{cheng_learned_2020}, which reduces the computation cost of the attention module originally proposed in \cite{wang_non-local_2018}.
The attention mechanism has been used in both \cite{cheng_learned_2020} and \cite{wang_non-local_2018} to improve the compression efficiency by focusing the neural network on regions in the image that require higher bit rate.
In our model, this will allow the model to allocate channel bandwidth and power resources optimally.
$g_\mathcal{C}$ refers to the two quantization strategies, which we will introduce next.

\subsection{Quantization}
\label{subs:quantization}

In order to produce an encoder that outputs channel symbols from a finite constellation, we perform quantization of the latent vector generated by the encoder, $\bar{\mathbf{z}}=g_\mathcal{C}(\mathbf{z})$. 
We will consider two quantization strategies: the \textit{soft-to-hard quantizer}, first introduced by \cite{Agustsson:softQuant:NIPS2017}, and the \textit{learned soft-to-hard quantizer}, which is our extension of the soft-to-hard quantizer that allows the constellation $\mathcal{C}$ to be learned as well.
We will first describe the soft-to-hard quantizer.

\subsubsection{Soft-to-hard quantizer}
\label{subsubsec:soft_hard_quantization}

\begin{figure}
\begin{center}
\includegraphics[width=0.7\linewidth]{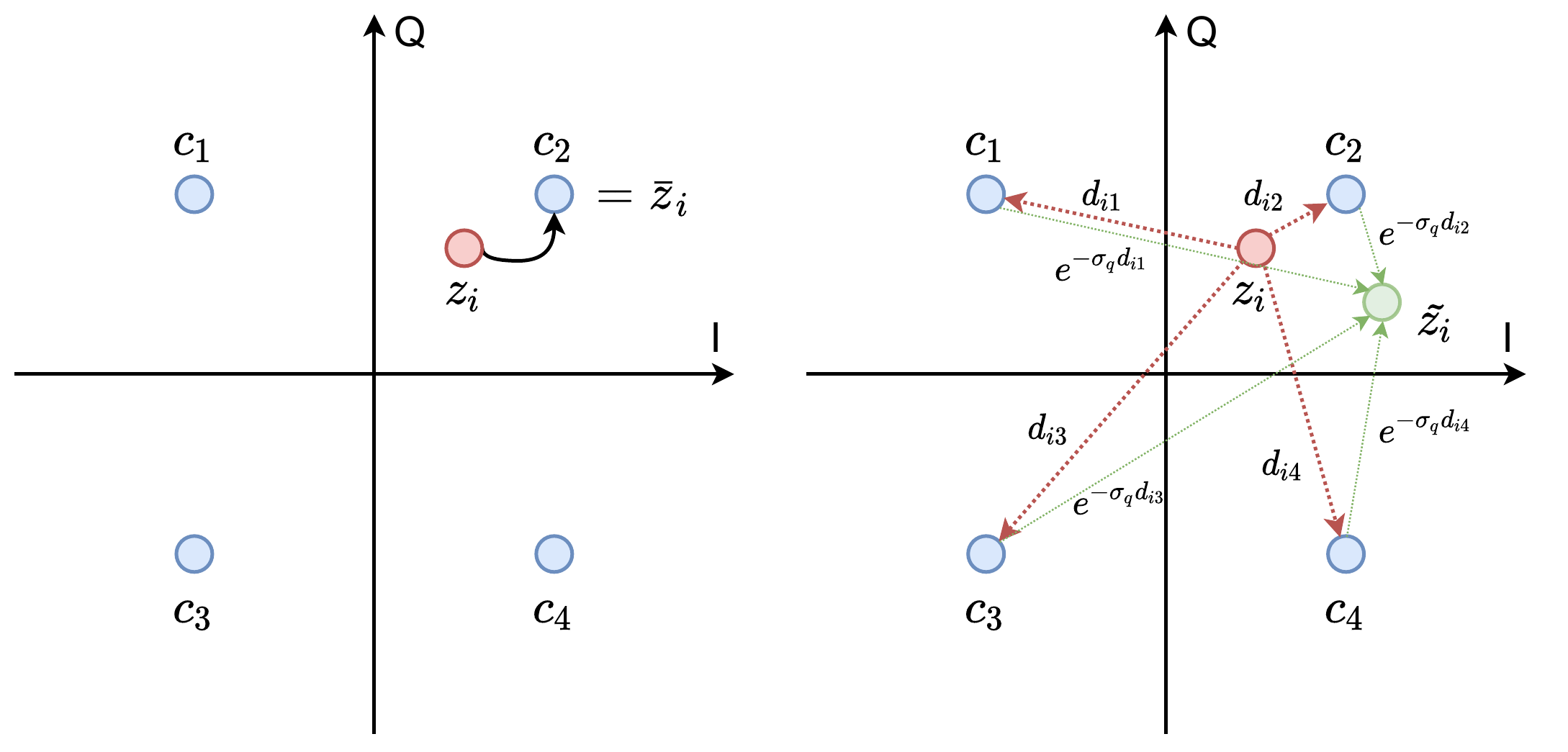}
\end{center}
  \caption{Illustration of the soft-to-hard quantization procedure for a single value $z_i$ using a QPSK constellation.
  The hard quatized value $\bar{z}_i$, on the left hand side, simply maps the latent value $z_i$ to the nearest point in the constellation, while the soft quantized value $\tilde{z}_i$ is the softmax weighted sum of the constellation points according to the squared $l_2$ distance of $z_i$ to each constellation point.}
\label{fig:soft_hard_quant}
\end{figure}

Given the encoder output $\mathbf{z}$, we first apply a ``hard" quantization, which simply maps element $z_i\in\mathbf{z}$ to the nearest symbol in $\mathcal{C}$.
This forms the channel input $\bar{\mathbf{z}}$.
However, this operation is not differentiable.
In order to obtain a differentiable approximation of the hard quantization operation, we will use the  ``soft" quantization approach, proposed in \cite{Agustsson:softQuant:NIPS2017}. 
In this approach, each quantized symbol is generated as the softmax weighted sum of the symbols in $\mathcal{C}$ based on their distances from $z_i$; 
that is,
\begin{equation}
\label{eq:soft_quantization}
    \tilde{z_i} = \sum_{j=1}^M \frac{e^{-\sigma_q d_{ij}}}{\sum_{n=1}^M e^{-\sigma_q d_{in}}} c_j,
\end{equation}
where $\sigma_q$ is a parameter controlling the ``hardness'' of the assignment, and $d_{ij} = ||z_i-c_j||^2_2$ is the squared $l_2$ distance between the latent value $z_i$ and the constellation point $c_j$.
As such, in the forward pass, the quantizer uses the hard quantization, corresponding to the channel input $\bar{\mathbf{z}}$, and in the backward pass, the gradient from the soft quantization $\tilde{\mathbf{z}}$ is used to update $\boldsymbol{\theta}$.
That is,
\begin{equation}
    \frac{\partial\bar{\mathbf{z}}}{\partial \mathbf{z}} =
    \frac{\partial\tilde{\mathbf{z}}}{\partial \mathbf{z}}.
\end{equation}
A diagram illustrating the soft-to-hard quantizer for 4-QAM, also known as the quadrature phase shift keying (QPSK), is shown in Fig. \ref{fig:soft_hard_quant}.

We consider constellation symbols that are uniformly distributed in a square lattice over the complex plane, similar to QAM-modulation. 
For QAM constellation consisting of $M$ symbols, denoted as M-QAM, we define the max amplitude $A_{max}$ and inter symbol distance $d_{sym}$ as:
\begin{equation}
    A_{max} = \frac{(M-1)}{2}\sqrt{\frac{12 \bar{P}}{(M^2-1)}},
\end{equation}
\begin{equation}
    d_{sym} = \sqrt{\frac{12 \bar{P}}{(M^2-1)}},
\end{equation}
where $\bar{P}$ is the average power of the constellation under uniform distribution, i.e., $\mathbb{E}[\mathcal{C}^2] = \frac{1}{M} \sum_{i=1}^M |{c}_{i}|^2 = \bar{P}$.

\subsubsection{Learned soft-to-hard quantizer}
\label{subsubsec:nonuniform}

For the learned soft-to-hard quantization, the process is the same except the constellation $\mathcal{C}$ is part of the parameters to be trained. 
That is, given Eqn. (\ref{eq:soft_quantization}), the gradient of the soft quantized value $\tilde{z}_i$ with respect to the input is
\begin{equation}
    \frac{\partial\tilde{z}_i}{\partial z_i} =
    \frac{\partial\mathcal{C}}{\partial z_i}
    \frac{\partial\tilde{z}_i}{\partial \mathcal{C}}.
\end{equation}
% and the constellation $\mathcal{C}$ is treated as part of the parameters and updated during training.

The constellation points are initialized in the same way, but are allowed to be updated during training.
In order to maintain the power constraint, we normalize the symbol power after each update
\begin{equation}
    c_i \leftarrow \frac{\sqrt{\bar{P}}}{\sqrt{\sum_{j=1}^M P(c_j) |c_j|^2}} c_i.
\end{equation}
Since we do not have the distribution over the constellation points $P(c_j),~j=1,...,M$, we estimate this probability distribution with a batch of input images $\{\mathbf{x}^v\}_{v=1}^B$, where $B$ is the batch size, utilizing the convex weights used in the soft assignment in Eq. (\ref{eq:soft_quantization}).
Since the weights sum to 1, we can treat them as probabilities and average the probability of each constellation point over the training batch to obtain an estimate of $P(c_j)$. 
That is, the probability of selecting a constellation point $c_j$ can be estimated as 
\begin{equation}
    \hat{P}(c_j) =
    \frac{1}{Bk}\sum_{v=1}^{B}\sum_{i=1}^{k}
    \frac{e^{-\sigma_q d_{ij}^{v}}}{\sum_{n=1}^M e^{-\sigma_q d_{in}^{v}}},
    \label{eq:constellation_likelihood}
\end{equation}
where $\hat{P}(c_j)$ is the empirical probability the constellation point $c_j$, estimated from a batch of input images $\{\mathbf{x}^v\}_{v=1}^B$, $d_{ij}^v=||z_i^v - c_j||_2^2$ is the squared $l_2$ distance between the $i$th element in the $v$th latent vector in the batch and the constellation point $c_j$.
The constellation symbol power is then normalized as
\begin{equation}
    c_i \leftarrow \frac{\sqrt{\bar{P}}}{\sqrt{\sum_{j=1}^M \hat{P}(c_j) |c_j|^2}} c_i.
\end{equation}

\subsection{Training Strategy}
\label{subsec:training}

In order to promote exploration of the available constellation points, we introduce a regularization term based on the Kullback-Leibler (KL) divergence between the distribution $P(\mathcal{C})$ and a uniform distribution over the constellation set $\mathcal{U}(\mathcal{C})$. 
The KL divergence between two distributions $P_W$ and $P_V$ is defined as
\begin{equation}
    D_{\text{KL}}(P_W\ ||\ P_V) =
    \mathbb{E}\left[\log\left(\frac{P_W}{P_V}\right)\right],
\end{equation}
and it measures how different the two distributions are, with $D_{\text{KL}}(P_W\ ||\ P_V)=0 \Longleftrightarrow P_W=P_V$. 
% The distribution $P(\mathcal{C})$ represents the probability of selecting a point in the constellation set $\mathcal{C}$ given the encoded latent vector $\mathbf{z}$.
By regularizing the distortion loss with the KL divergence $D_{\text{KL}}(P(\mathcal{C})\ ||\ \mathcal{U}(\mathcal{C}))$, we encourage the quantizer $q_\mathcal{C}$ to explore the available constellation points, which may improve the end-to-end performance of the system.
The distribution $P(\mathcal{C})$ is estimated as in Eqn. (\ref{eq:constellation_likelihood}). 
Therefore, the final loss function we use for training is:
\begin{equation}
\label{eq:loss}
    l(\mathbf{x},\hat{\mathbf{x}}) = d(\mathbf{x}, \hat{\mathbf{x}}) + \lambda D_{\text{KL}}(\hat{P}(\mathcal{C})\ ||\ \mathcal{U}(\mathcal{C})),
\end{equation}
where $d(\cdot, \cdot)$ is the distortion measure (MSE if evaluating on the PSNR metric or 1-MS-SSIM if evaluating on the MS-SSIM metric) and $\lambda$ is the weighting parameter to control the amount of regularization.

\section{Experimental Results}
\label{sec:results}

% \subsection{Numerical results}
% \label{subsec:numerical_results}

Herein, we perform a series of experiments to demonstrate the performance of \emph{DeepJSCC-Q}. 
For the first CSI acquisition scenario, defined in Sec. \ref{sec:problem_def}, we will consider a constant channel gain magnitude $|h| = 1$, which implies a static AWGN channel (referred to as ``AWGN" channel henceforth), while in the second scenario, we consider $h \sim CN(0,1)$ and will refer to it as the ``slow fading" channel.
We train the model for $\text{SNR}_{\text{Train}} \in \{7, 10, 16\}$dB on the ImageNet dataset \cite{deng2009imagenet} which consists of 1.2 million RGB images of various resolution. 
We split the dataset into $9:1$ for training and validation, respectively.
In order to train in batches, we take random crops of $128 \times 128$ from the training images.
For final evaluation, we use the Kodak dataset\footnote{http://r0k.us/graphics/kodak/} consisting of 24 $768 \times 512$ images.
We use the Pytorch \cite{paszke_automatic_2017} library and the Adam \cite{kingma_adam_2017} optimizer with $\beta_1=0.9$, and $\beta_2=0.99$ to train our encoder and decoder networks.
The learning rate was initialized at $0.0001$ for the AWGN channel case, while for the slow fading case, we initialized at $0.00005$.
We use a batch size of 32 and early stopping with a patience of 8 epochs, where the maximum number of training epochs is 1000.
We implement learning rate scheduling, where the learning rate is reduced by a factor of $0.8$ if the loss does not improve for 4 epochs consecutively.
We use an average constellation power $\bar{P} = 1$ and change the channel noise power $\sigma^2$ accordingly to obtain any given SNR. 
The soft-to-hard hardness parameter $\sigma_q$ is linearly annealed using the annealing function
\begin{equation}
    \sigma_q^{(t)} = \min \left(100, \sigma_q^{(t-1)} + 5 \left\lfloor \frac{t}{10000} \right\rfloor \right),
\end{equation}
where $t$ is the parameter update step number and we initialize $\sigma_q^{(0)} = 5$.
We set the weighting for the regularizer $\lambda=0.05$ for the AWGN channel case when the size of the constellation is relatively small, i.e., $M < 4096$, as we empirically found it to be helpful to encourage the channel input to be more uniformly distributed across the constellation set, while for $M \geq 4096$, $\lambda=0$ performed better, indicating that it is more beneficial to choose a subset of available symbols with higher probability than using all symbols with the same frequency.
For the slow fading channel case, we found $\lambda=0$ to produce better results regardless of the constellation order. 
% The result of the training using the aforementioned parameters are shown in Fig. \ref{fig:training_result} for both soft-to-hard and learned soft-to-hard quantization for the AWGN channel.

In order to compare the performance of our solution, we consider a baseline separation scheme, in which images are first compressed into bits using the BPG image compression codec \cite{Bellard:BPG} and then protected from channel distortion with low density parity check (LDPC) codes.
The LDPC codes we use are from the IEEE 802.11ad standard \cite{noauthor_ieee_2012}, with block length 672 bits for both rate $1/2$ and $3/4$ codes.
We will compare the average image quality over the evaluation dataset, with error bars showing the standard deviation of the image quality metric across the dataset.
We refer to M-ary constellations using soft-to-hard quantization as ``M-QAM", indicating the constellation is a standard square QAM, whereas for the learned constellations, we refer to them as ``L-M".

\begin{figure} 
  \centering
  \subfloat[PSNR\label{subfig:psnr_graceful}]{%
    \begin{tikzpicture}
        \pgfplotsset{
            legend style={
                font=\fontsize{4}{4}\selectfont,
                at={(1.0,.0)},
                anchor=south east,
            },
            width=0.5\textwidth,
            xmin=0,
            xmax=17,
            ymin=20,
            ymax=44,
            xtick distance=2,
            ytick distance=2,
            xlabel={SNR (dB)},
            ylabel={PSNR (dB)},
            grid=both,
            grid style={line width=.1pt, draw=gray!10},
            major grid style={line width=.2pt,draw=gray!50},
            every axis/.append style={
                x label style={
                    font=\fontsize{8}{8}\selectfont,
                    at={(axis description cs:0.5,-0.04)},
                    },
                y label style={
                    font=\fontsize{8}{8}\selectfont,
                    at={(axis description cs:-0.08,0.5)},
                    },
                x tick label style={
                    font=\fontsize{8}{8}\selectfont,
                    /pgf/number format/.cd,
                    fixed,
                    fixed zerofill,
                    precision=0,
                    /tikz/.cd
                    },
                y tick label style={
                    font=\fontsize{8}{8}\selectfont,
                    /pgf/number format/.cd,
                    fixed,
                    fixed zerofill,
                    precision=1,
                    /tikz/.cd
                    },
            }
        }
        \begin{axis}
        \addplot[blue, solid, line width=0.9pt, 
        mark=*, mark options={fill=blue, scale=1.1}, 
        error bars/.cd, y dir=both, y explicit, every nth mark=2] 
                table [x=snr, y=soft_hard, 
                y error=soft_hard_std, col sep=comma]
                {data/djsccq_4096qam_c256_tsnr16_psnr.csv};
        \addlegendentry{\textit{DeepJSCC-Q} 4096-QAM 
        ($\lambda=0$, $\text{SNR}_{\text{Train}}=16dB$)}
        
        \addplot[magenta, solid, line width=0.9pt, 
        mark=*, mark options={fill=magenta, scale=1.1}, 
        error bars/.cd, y dir=both, y explicit, every nth mark=2] 
                table [x=snr, y=soft_hard_klloss, 
                y error=soft_hard_klloss_std, col sep=comma]
                {data/djsccq_64qam_c256_tsnr10_psnr.csv};
        \addlegendentry{\textit{DeepJSCC-Q} 64-QAM 
        ($\lambda=0.05$, $\text{SNR}_{\text{Train}}=10dB$)}
        
        % \addplot[orange, solid, line width=0.9pt, 
        % mark=*, mark options={fill=orange, scale=1.1}, 
        % error bars/.cd, y dir=both, y explicit, every nth mark=3] 
        %         table [x=snr, y=nonuniform, 
        %         y error=nonuniform_std, col sep=comma]
        %         {data/djsccq_64qam_c256_tsnr10_psnr.csv};
        % \addlegendentry{\textit{DeepJSCC-Q} L-64 
        % ($\lambda=0.05$, $\text{SNR}_{\text{Train}}=10dB$)}
        
        \addplot[green, solid, line width=0.9pt, 
        mark=*, mark options={fill=green, scale=1.1}, 
        error bars/.cd, y dir=both, y explicit, every nth mark=2
        ] 
        table [x=snr, y=soft_hard_klloss, 
                y error=soft_hard_klloss_std, col sep=comma]
                {data/djsccq_16qam_c256_tsnr7_psnr.csv};
        \addlegendentry{\textit{DeepJSCC-Q} 16-QAM 
        ($\lambda=0.05$, $\text{SNR}_{\text{Train}}=7dB$)}
        
        \addplot[color=black, solid, line width=1.2pt, 
        mark=*, mark options={fill=black, solid, scale=1.1}, 
        error bars/.cd, y dir=both, y explicit, every nth mark=2,
        error bar style={mark=*, line width=1pt},
        error mark options={rotate=90, black, mark size=2pt, line width=4pt}
        ] 
        table [x=snr, y=bpsk_05, 
        y error=bpsk_05_std, col sep=comma]
        {data/bpg_c256_psnr.csv};
        \addlegendentry{BPG + LDPC 1/2 BPSK}
        
        % \addplot[color=red, dashed, line width=1.2pt, 
        % mark=triangle*, mark options={fill=red, solid, scale=1.1},
        % error bars/.cd, y dir=both, y explicit, every nth mark=4,
        % error bar style={mark=*, line width=2pt},
        % error mark options={rotate=90, red, mark size=2pt, line width=2pt}
        % ] 
        % table [x=snr, y=bpsk_075, 
        % y error=bpsk_075_std, col sep=comma]
        % {data/bpg_c256_psnr.csv};
        % \addlegendentry{BPG + LDPC 3/4 BPSK}
        
        \addplot[color=brown, solid, line width=1.2pt, 
        mark=square*, mark options={fill=brown, solid, scale=1.1},
        error bars/.cd, y dir=both, y explicit, every nth mark=2,
        error bar style={mark=*, line width=2pt},
        error mark options={rotate=90, brown, mark size=2pt, line width=6pt}
        ] 
        table [x=snr, y=qpsk_05, 
        y error=qpsk_05_std, col sep=comma]
        {data/bpg_c256_psnr.csv};
        \addlegendentry{BPG + LDPC 1/2 QPSK}
        
        \addplot[color=red, solid, line width=1.2pt, 
        mark=square*, mark options={fill=red, solid, scale=1.1},
        error bars/.cd, y dir=both, y explicit, every nth mark=2,
        error bar style={mark=*, line width=2pt},
        error mark options={rotate=90, red, mark size=2pt, line width=6pt}
        ] 
        table [x=snr, y=qpsk_075, 
        y error=qpsk_075_std, col sep=comma]
        {data/bpg_c256_psnr.csv};
        \addlegendentry{BPG + LDPC 3/4 QPSK}
        
        \addplot[color=teal, solid, line width=1.2pt, 
        mark=*, mark options={fill=teal, solid, scale=1.1},
        error bars/.cd, y dir=both, y explicit, every nth mark=2,
        error bar style={mark=*, line width=2pt},
        error mark options={rotate=90, teal, mark size=2pt, line width=6pt}
        ] 
        table [x=snr, y=16qam_05, 
        y error=16qam_05_std, col sep=comma]
        {data/bpg_c256_psnr.csv};
        \addlegendentry{BPG + LDPC 1/2 16-QAM}
        \end{axis}
        \end{tikzpicture}
    }
    \hfill
  \subfloat[MS-SSIM\label{subfig:msssim_graceful}]{%
    \begin{tikzpicture}
        \pgfplotsset{
            legend style={
                font=\fontsize{4}{4}\selectfont,
                at={(1.0,0.)},
                anchor=south east,
            },
            % height=0.4\linewidth,
            width=0.5\textwidth,
            xmin=0,
            xmax=17,
            ymin=0.9,
            ymax=1,
            xtick distance=2,
            ytick distance=0.02,
            xlabel={SNR (dB)},
            ylabel={MS-SSIM},
            grid=both,
            grid style={line width=.1pt, draw=gray!10},
            major grid style={line width=.2pt,draw=gray!50},
            every axis/.append style={
                x label style={
                    font=\fontsize{8}{8}\selectfont,
                    at={(axis description cs:0.5, -0.04)},
                    },
                y label style={
                    font=\fontsize{8}{8}\selectfont,
                    at={(axis description cs:-0.1,0.5)},
                    },
                x tick label style={
                    font=\fontsize{8}{8}\selectfont,
                    /pgf/number format/.cd,
                    fixed,
                    fixed zerofill,
                    precision=0,
                    /tikz/.cd
                    },
                y tick label style={
                    font=\fontsize{8}{8}\selectfont,
                    /pgf/number format/.cd,
                    fixed,
                    fixed zerofill,
                    precision=2,
                    /tikz/.cd
                    },
            }
        }
        \begin{axis}%[mark options={solid}]
        \addplot[blue, solid, line width=0.9pt, 
        mark=*, mark options={fill=blue, scale=1.1}, 
        error bars/.cd, y dir=both, y explicit, every nth mark=2] 
                table [x=snr, y=soft_hard, 
                y error=soft_hard_std, col sep=comma]
                {data/djsccq_4096qam_c256_tsnr16_msssim.csv};
        \addlegendentry{\textit{DeepJSCC-Q} 4096-QAM 
        ($\lambda=0$, $\text{SNR}_{\text{Train}}=16dB$)}
        
        \addplot[magenta, solid, line width=0.9pt, 
        mark=*, mark options={fill=magenta, scale=1.1}, 
        error bars/.cd, y dir=both, y explicit, every nth mark=2] 
                table [x=snr, y=soft_hard_klloss, 
                y error=soft_hard_klloss_std, col sep=comma]
                {data/djsccq_64qam_c256_tsnr10_msssim.csv};
        \addlegendentry{\textit{DeepJSCC-Q} 64-QAM 
        ($\lambda=0.05$, $\text{SNR}_{\text{Train}}=10dB$)}
        
        % \addplot[orange, solid, line width=0.9pt, 
        % mark=*, mark options={fill=orange, scale=1.1}, error bars/.cd, y dir=both, y explicit, every nth mark=1] 
        %         table [x=snr, y=nonuniform, 
        %         y error=soft_hard_klloss_std, col sep=comma]
        %         {data/djsccq_64qam_c256_tsnr10_msssim.csv};
        % \addlegendentry{\textit{DeepJSCC-Q} L-64 
        % ($\lambda=0.05$, $\text{SNR}_{\text{Train}}=10dB$)}
        
        \addplot[green, solid, line width=0.9pt, 
        mark=*, mark options={fill=green, scale=1.1}, 
        error bars/.cd, y dir=both, y explicit, every nth mark=2] 
                table [x=snr, y=soft_hard_klloss, 
                y error=soft_hard_klloss_std, col sep=comma]
                {data/djsccq_16qam_c256_tsnr7_msssim.csv};
        \addlegendentry{\textit{DeepJSCC-Q} 16-QAM 
        ($\lambda=0.05$, $\text{SNR}_{\text{Train}}=7dB$)}
        
        \addplot[color=black, solid, line width=1.2pt, 
        mark=*, mark options={fill=black, solid, scale=1.1}, 
        error bars/.cd, y dir=both, y explicit, every nth mark=2,
        error bar style={mark=*, line width=1pt},
        error mark options={rotate=90, black, mark size=2pt, line width=4pt}
        ] 
        table [x=snr, y=bpsk_05, 
        y error=bpsk_05_std, col sep=comma]
        {data/bpg_c256_msssim.csv};
        \addlegendentry{BPG + LDPC 1/2 BPSK}
        
        % \addplot[color=red, dashed, line width=1.2pt, 
        % mark=triangle*, mark options={fill=red, solid, scale=1.1},
        % error bars/.cd, y dir=both, y explicit, every nth mark=4,
        % error bar style={mark=*, line width=2pt},
        % error mark options={rotate=90, red, mark size=2pt, line width=2pt}
        % ] 
        % table [x=snr, y=bpsk_075, 
        % y error=bpsk_075_std, col sep=comma]
        % {data/bpg_c256_psnr.csv};
        % \addlegendentry{BPG + LDPC 3/4 BPSK}
        
        \addplot[color=brown, solid, line width=1.2pt, 
        mark=square*, mark options={fill=brown, solid, scale=1.1},
        error bars/.cd, y dir=both, y explicit, every nth mark=2,
        error bar style={mark=*, line width=2pt},
        error mark options={rotate=90, brown, mark size=2pt, line width=6pt}
        ] 
        table [x=snr, y=qpsk_05, 
        y error=qpsk_05_std, col sep=comma]
        {data/bpg_c256_msssim.csv};
        \addlegendentry{BPG + LDPC 1/2 QPSK}
        
        \addplot[color=red, solid, line width=1.2pt, 
        mark=square*, mark options={fill=red, solid, scale=1.1},
        error bars/.cd, y dir=both, y explicit, every nth mark=2,
        error bar style={mark=*, line width=2pt},
        error mark options={rotate=90, red, mark size=2pt, line width=6pt}
        ] 
        table [x=snr, y=qpsk_075, 
        y error=qpsk_075_std, col sep=comma]
        {data/bpg_c256_msssim.csv};
        \addlegendentry{BPG + LDPC 3/4 QPSK}
        
        \addplot[color=teal, solid, line width=1.2pt, 
        mark=*, mark options={fill=teal, solid, scale=1.1},
        error bars/.cd, y dir=both, y explicit, every nth mark=2,
        error bar style={mark=*, line width=2pt},
        error mark options={rotate=90, teal, mark size=2pt, line width=6pt}
        ] 
        table [x=snr, y=16qam_05, 
        y error=16qam_05_std, col sep=comma]
        {data/bpg_c256_msssim.csv};
        \addlegendentry{BPG + LDPC 1/2 16-QAM}
        \end{axis}
        \end{tikzpicture}
        }
  \caption{
  Comparison of \emph{DeepJSCC-Q} trained for different $\text{SNR}_{\text{Train}}$ with the separation approach using BPG for source coding and LDPC codes for channel coding in the AWGN channel case.
  }
  \label{fig:graceful} 
\end{figure}
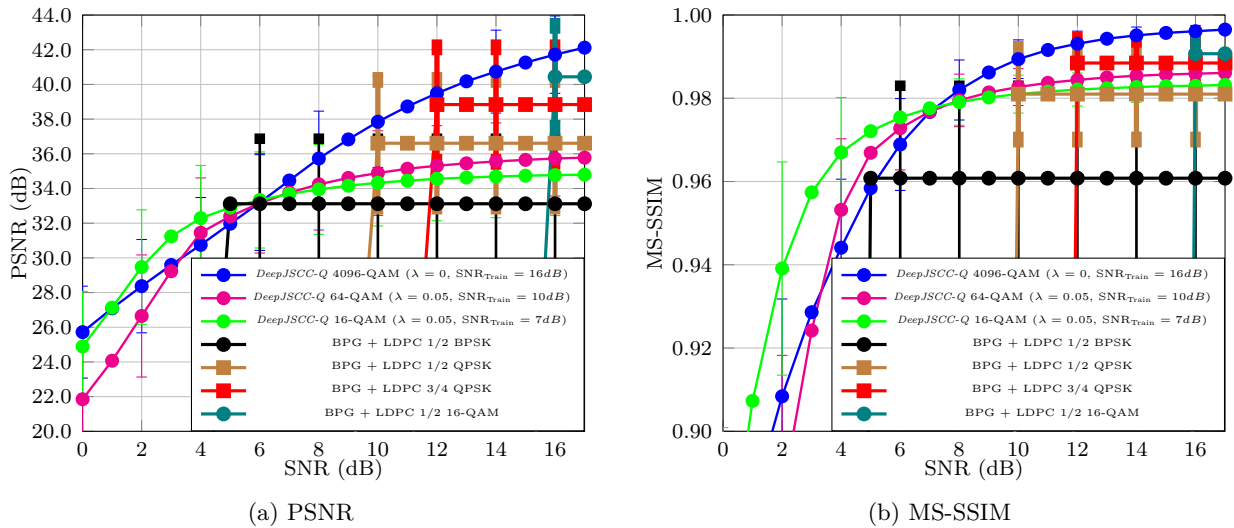

\begin{figure} 
  \centering
  \subfloat[PSNR\label{subfig:psnr_graceful_4096}]{%
    \begin{tikzpicture}
        \pgfplotsset{
            legend style={
                font=\fontsize{4}{4}\selectfont,
                at={(1.0,.0)},
                anchor=south east,
            },
            width=0.5\linewidth,
            xmin=0,
            xmax=17,
            ymin=20,
            ymax=44,
            xtick distance=2,
            ytick distance=2,
            xlabel={SNR (dB)},
            ylabel={PSNR (dB)},
            grid=both,
            grid style={line width=.1pt, draw=gray!10},
            major grid style={line width=.2pt,draw=gray!50},
            every axis/.append style={
                x label style={
                    font=\fontsize{8}{8}\selectfont,
                    at={(axis description cs:0.5,-0.04)},
                    },
                y label style={
                    font=\fontsize{8}{8}\selectfont,
                    at={(axis description cs:-0.08,0.5)},
                    },
                x tick label style={
                    font=\fontsize{8}{8}\selectfont,
                    /pgf/number format/.cd,
                    fixed,
                    fixed zerofill,
                    precision=0,
                    /tikz/.cd
                    },
                y tick label style={
                    font=\fontsize{8}{8}\selectfont,
                    /pgf/number format/.cd,
                    fixed,
                    fixed zerofill,
                    precision=1,
                    /tikz/.cd
                    },
            }
        }
        \begin{axis}
        \addplot[blue, solid, line width=0.9pt, 
        mark=*, mark options={fill=blue, scale=1.1}, 
        error bars/.cd, y dir=both, y explicit, every nth mark=2] 
                table [x=snr, y=soft_hard, 
                y error=soft_hard_std, col sep=comma]
                {data/djsccq_4096qam_c256_tsnr16_psnr.csv};
        \addlegendentry{\textit{DeepJSCC-Q} 4096-QAM 
        ($\lambda=0$, $\text{SNR}_{\text{Train}}=16dB$)}
        
        \addplot[magenta, solid, line width=0.9pt, 
        mark=*, mark options={fill=magenta, scale=1.1}, 
        error bars/.cd, y dir=both, y explicit, every nth mark=2] 
                table [x=snr, y=soft_hard, 
                y error=soft_hard_std, col sep=comma]
                {data/djsccq_4096qam_c256_tsnr10_psnr.csv};
        \addlegendentry{\textit{DeepJSCC-Q} 4096-QAM 
        ($\lambda=0.05$, $\text{SNR}_{\text{Train}}=10dB$)}
        
        \addplot[green, solid, line width=0.9pt, 
        mark=*, mark options={fill=green, scale=1.1}, 
        error bars/.cd, y dir=both, y explicit, every nth mark=2
        ] 
        table [x=snr, y=soft_hard, 
                y error=soft_hard_std, col sep=comma]
                {data/djsccq_4096qam_c256_tsnr7_psnr.csv};
        \addlegendentry{\textit{DeepJSCC-Q} 4096-QAM 
        ($\lambda=0.05$, $\text{SNR}_{\text{Train}}=7dB$)}
        
        \addplot[color=black, solid, line width=1.2pt, 
        mark=*, mark options={fill=black, solid, scale=1.1}, 
        error bars/.cd, y dir=both, y explicit, every nth mark=2,
        error bar style={mark=*, line width=1pt},
        error mark options={rotate=90, black, mark size=2pt, line width=4pt}
        ] 
        table [x=snr, y=bpsk_05, 
        y error=bpsk_05_std, col sep=comma]
        {data/bpg_c256_psnr.csv};
        \addlegendentry{BPG + LDPC 1/2 BPSK}
        
        % \addplot[color=red, dashed, line width=1.2pt, 
        % mark=triangle*, mark options={fill=red, solid, scale=1.1},
        % error bars/.cd, y dir=both, y explicit, every nth mark=4,
        % error bar style={mark=*, line width=2pt},
        % error mark options={rotate=90, red, mark size=2pt, line width=2pt}
        % ] 
        % table [x=snr, y=bpsk_075, 
        % y error=bpsk_075_std, col sep=comma]
        % {data/bpg_c256_psnr.csv};
        % \addlegendentry{BPG + LDPC 3/4 BPSK}
        
        \addplot[color=brown, solid, line width=1.2pt, 
        mark=square*, mark options={fill=brown, solid, scale=1.1},
        error bars/.cd, y dir=both, y explicit, every nth mark=2,
        error bar style={mark=*, line width=2pt},
        error mark options={rotate=90, brown, mark size=2pt, line width=6pt}
        ] 
        table [x=snr, y=qpsk_05, 
        y error=qpsk_05_std, col sep=comma]
        {data/bpg_c256_psnr.csv};
        \addlegendentry{BPG + LDPC 1/2 QPSK}
        
        \addplot[color=red, solid, line width=1.2pt, 
        mark=square*, mark options={fill=red, solid, scale=1.1},
        error bars/.cd, y dir=both, y explicit, every nth mark=2,
        error bar style={mark=*, line width=2pt},
        error mark options={rotate=90, red, mark size=2pt, line width=6pt}
        ] 
        table [x=snr, y=qpsk_075, 
        y error=qpsk_075_std, col sep=comma]
        {data/bpg_c256_psnr.csv};
        \addlegendentry{BPG + LDPC 3/4 QPSK}
        
        \addplot[color=teal, solid, line width=1.2pt, 
        mark=*, mark options={fill=teal, solid, scale=1.1},
        error bars/.cd, y dir=both, y explicit, every nth mark=2,
        error bar style={mark=*, line width=2pt},
        error mark options={rotate=90, teal, mark size=2pt, line width=6pt}
        ] 
        table [x=snr, y=16qam_05, 
        y error=16qam_05_std, col sep=comma]
        {data/bpg_c256_psnr.csv};
        \addlegendentry{BPG + LDPC 1/2 16-QAM}
        \end{axis}
        \end{tikzpicture}
    }
    \hfill
  \subfloat[MS-SSIM\label{subfig:msssim_graceful_4096}]{%
    \begin{tikzpicture}
        \pgfplotsset{
            legend style={
                font=\fontsize{4}{4}\selectfont,
                at={(1.0,0.)},
                anchor=south east,
            },
            % height=0.4\linewidth,
            width=0.5\textwidth,
            xmin=0,
            xmax=17,
            ymin=0.9,
            ymax=1,
            xtick distance=2,
            ytick distance=0.02,
            xlabel={SNR (dB)},
            ylabel={MS-SSIM},
            grid=both,
            grid style={line width=.1pt, draw=gray!10},
            major grid style={line width=.2pt,draw=gray!50},
            every axis/.append style={
                x label style={
                    font=\fontsize{8}{8}\selectfont,
                    at={(axis description cs:0.5, -0.04)},
                    },
                y label style={
                    font=\fontsize{8}{8}\selectfont,
                    at={(axis description cs:-0.1,0.5)},
                    },
                x tick label style={
                    font=\fontsize{8}{8}\selectfont,
                    /pgf/number format/.cd,
                    fixed,
                    fixed zerofill,
                    precision=0,
                    /tikz/.cd
                    },
                y tick label style={
                    font=\fontsize{8}{8}\selectfont,
                    /pgf/number format/.cd,
                    fixed,
                    fixed zerofill,
                    precision=2,
                    /tikz/.cd
                    },
            }
        }
        \begin{axis}%[mark options={solid}]
        \addplot[blue, solid, line width=0.9pt, 
        mark=*, mark options={fill=blue, scale=1.1}, 
        error bars/.cd, y dir=both, y explicit, every nth mark=2] 
                table [x=snr, y=soft_hard, 
                y error=soft_hard_std, col sep=comma]
                {data/djsccq_4096qam_c256_tsnr16_msssim.csv};
        \addlegendentry{\textit{DeepJSCC-Q} 4096-QAM 
        ($\lambda=0$, $\text{SNR}_{\text{Train}}=16dB$)}
        
        \addplot[magenta, solid, line width=0.9pt, 
        mark=*, mark options={fill=magenta, scale=1.1}, 
        error bars/.cd, y dir=both, y explicit, every nth mark=2] 
                table [x=snr, y=soft_hard, 
                y error=soft_hard_std, col sep=comma]
                {data/djsccq_4096qam_c256_tsnr10_msssim.csv};
        \addlegendentry{\textit{DeepJSCC-Q} 4096-QAM 
        ($\lambda=0.05$, $\text{SNR}_{\text{Train}}=10dB$)}
        
        % \addplot[orange, solid, line width=0.9pt, 
        % mark=*, mark options={fill=orange, scale=1.1}, error bars/.cd, y dir=both, y explicit, every nth mark=1] 
        %         table [x=snr, y=nonuniform, 
        %         y error=soft_hard_klloss_std, col sep=comma]
        %         {data/djsccq_64qam_c256_tsnr10_msssim.csv};
        % \addlegendentry{\textit{DeepJSCC-Q} L-64 
        % ($\lambda=0.05$, $\text{SNR}_{\text{Train}}=10dB$)}
        
        \addplot[green, solid, line width=0.9pt, 
        mark=*, mark options={fill=green, scale=1.1}, 
        error bars/.cd, y dir=both, y explicit, every nth mark=2] 
                table [x=snr, y=soft_hard, 
                y error=soft_hard_std, col sep=comma]
                {data/djsccq_4096qam_c256_tsnr7_msssim.csv};
        \addlegendentry{\textit{DeepJSCC-Q} 4096-QAM 
        ($\lambda=0.05$, $\text{SNR}_{\text{Train}}=7dB$)}
        
        \addplot[color=black, solid, line width=1.2pt, 
        mark=*, mark options={fill=black, solid, scale=1.1}, 
        error bars/.cd, y dir=both, y explicit, every nth mark=2,
        error bar style={mark=*, line width=1pt},
        error mark options={rotate=90, black, mark size=2pt, line width=4pt}
        ] 
        table [x=snr, y=bpsk_05, 
        y error=bpsk_05_std, col sep=comma]
        {data/bpg_c256_msssim.csv};
        \addlegendentry{BPG + LDPC 1/2 BPSK}
        
        % \addplot[color=red, dashed, line width=1.2pt, 
        % mark=triangle*, mark options={fill=red, solid, scale=1.1},
        % error bars/.cd, y dir=both, y explicit, every nth mark=4,
        % error bar style={mark=*, line width=2pt},
        % error mark options={rotate=90, red, mark size=2pt, line width=2pt}
        % ] 
        % table [x=snr, y=bpsk_075, 
        % y error=bpsk_075_std, col sep=comma]
        % {data/bpg_c256_psnr.csv};
        % \addlegendentry{BPG + LDPC 3/4 BPSK}
        
        \addplot[color=brown, solid, line width=1.2pt, 
        mark=square*, mark options={fill=brown, solid, scale=1.1},
        error bars/.cd, y dir=both, y explicit, every nth mark=2,
        error bar style={mark=*, line width=2pt},
        error mark options={rotate=90, brown, mark size=2pt, line width=6pt}
        ] 
        table [x=snr, y=qpsk_05, 
        y error=qpsk_05_std, col sep=comma]
        {data/bpg_c256_msssim.csv};
        \addlegendentry{BPG + LDPC 1/2 QPSK}
        
        \addplot[color=red, solid, line width=1.2pt, 
        mark=square*, mark options={fill=red, solid, scale=1.1},
        error bars/.cd, y dir=both, y explicit, every nth mark=2,
        error bar style={mark=*, line width=2pt},
        error mark options={rotate=90, red, mark size=2pt, line width=6pt}
        ] 
        table [x=snr, y=qpsk_075, 
        y error=qpsk_075_std, col sep=comma]
        {data/bpg_c256_msssim.csv};
        \addlegendentry{BPG + LDPC 3/4 QPSK}
        
        \addplot[color=teal, solid, line width=1.2pt, 
        mark=*, mark options={fill=teal, solid, scale=1.1},
        error bars/.cd, y dir=both, y explicit, every nth mark=2,
        error bar style={mark=*, line width=2pt},
        error mark options={rotate=90, teal, mark size=2pt, line width=6pt}
        ] 
        table [x=snr, y=16qam_05, 
        y error=16qam_05_std, col sep=comma]
        {data/bpg_c256_msssim.csv};
        \addlegendentry{BPG + LDPC 1/2 16-QAM}
        \end{axis}
        \end{tikzpicture}
        }
  \caption{
  Comparison of \emph{DeepJSCC-Q} using modulation order $M = 4096$ trained for different $\text{SNR}_{\text{Train}}$ with the separation approach using BPG for source coding and LDPC codes for channel coding in the AWGN channel case.
  }
  \label{fig:graceful_4096} 
\end{figure}
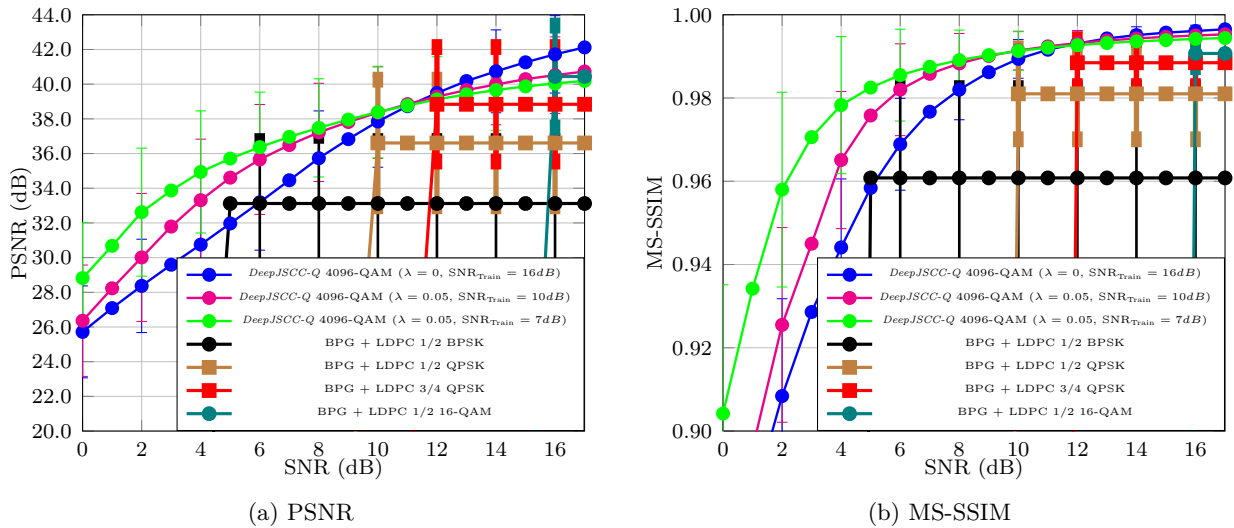

Fig. \ref{fig:graceful} shows the result of \emph{DeepJSCC-Q} trained for different $\text{SNR}_{\text{Train}}$ and modulation order $M$ and tested over a range of channel SNR values for the AWGN channel case.
For all M-QAM constellations shown here, \emph{DeepJSCC-Q} exhibited graceful degradation of image quality with decreasing channel quality.
This is similar to the DeepJSCC results in \cite{Eirina:TCCN:19}, but we are able to obtain the same behavior despite being constrained to a finite constellation.
Moreover, when compared with the separation-based results, the \emph{DeepJSCC-Q} 4096-QAM model performed almost exactly on the envelope of all the separation-based schemes.
Although the \emph{DeepJSCC-Q} models trained with modulation orders $M=16,64$ did not perform as well as the separation-based schemes, increasing the modulation order at those SNRs can improve the performance of \emph{DeepJSCC-Q}, as shown in Fig. \ref{fig:graceful_4096}.
We see that when we employ a modulation order of $M=4096$ for $\text{SNR}_{\text{Train}} = 7, 10$dB, \emph{DeepJSCC-Q} beats the separation-based schemes convincingly.
This shows that, the end-to-end optimization of \emph{DeepJSCC-Q} is fundamentally different from separation-based schemes, as the end-to-end distortion is directly affected by the channel distortion rather than the successful decoding of the channel code for a given source code rate.

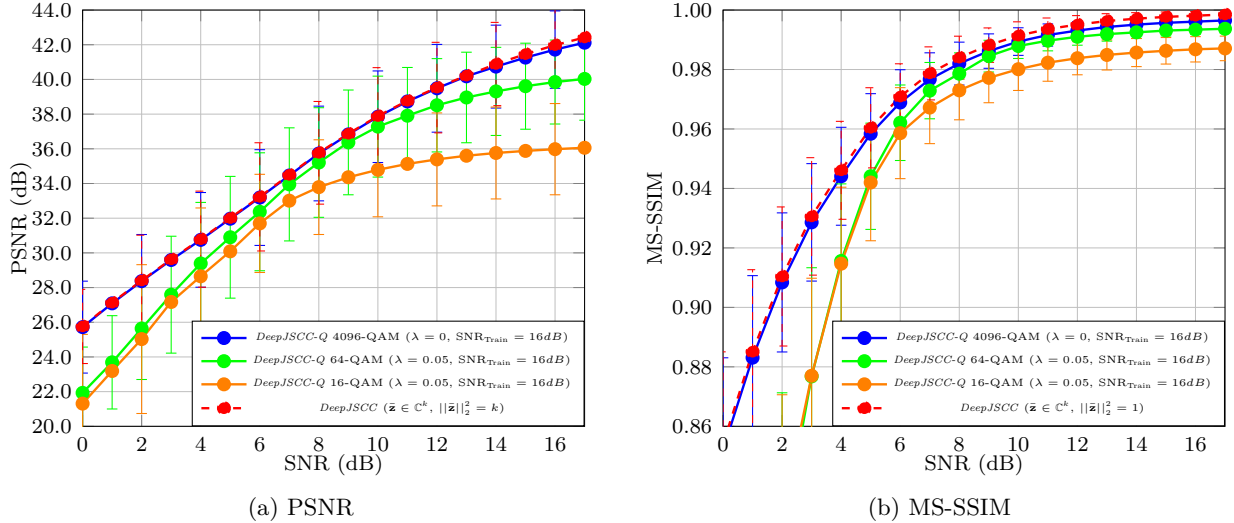
\begin{figure} 
    \centering
  \subfloat[PSNR\label{subfig:psnr_qam_v_noquant}]{%
    \begin{tikzpicture}
        \pgfplotsset{
            legend style={
                font=\fontsize{4}{4}\selectfont,
                at={(1.0,.0)},
                anchor=south east,
            },
            width=0.5\linewidth,
            xmin=0,
            xmax=17,
            ymin=20,
            ymax=44,
            xtick distance=2,
            ytick distance=2,
            xlabel={SNR (dB)},
            ylabel={PSNR (dB)},
            grid=both,
            grid style={line width=.1pt, draw=gray!10},
            major grid style={line width=.2pt,draw=gray!50},
            every axis/.append style={
                x label style={
                    font=\fontsize{8}{8}\selectfont,
                    at={(axis description cs:0.5,-0.04)},
                    },
                y label style={
                    font=\fontsize{8}{8}\selectfont,
                    at={(axis description cs:-0.08,0.5)},
                    },
                x tick label style={
                    font=\fontsize{8}{8}\selectfont,
                    /pgf/number format/.cd,
                    fixed,
                    fixed zerofill,
                    precision=0,
                    /tikz/.cd
                    },
                y tick label style={
                    font=\fontsize{8}{8}\selectfont,
                    /pgf/number format/.cd,
                    fixed,
                    fixed zerofill,
                    precision=1,
                    /tikz/.cd
                    },
            }
        }
        \begin{axis}
        \addplot[blue, solid, line width=0.9pt, 
        mark=*, mark options={fill=blue, scale=1.1}, 
        error bars/.cd, y dir=both, y explicit, every nth mark=2] 
                table [x=snr, y=soft_hard, 
                y error=soft_hard_std, col sep=comma]
                {data/djsccq_4096qam_c256_tsnr16_psnr.csv};
        \addlegendentry{\textit{DeepJSCC-Q} 4096-QAM 
        ($\lambda=0$, $\text{SNR}_{\text{Train}}=16dB$)}
        
        % \addplot[orange, dashed, line width=0.9pt, 
        % mark=*, mark options={fill=orange, scale=1.1}, error bars/.cd, y dir=both, y explicit, every nth mark=1] 
        %         table [x=snr, y=soft_hard_klloss,
        %         y error=soft_hard_klloss_std, col sep=comma]
        %         {data/djsccq_1024qam_c256_tsnr16_psnr.csv};
        % \addlegendentry{\textit{DeepJSCC-Q} 1024-QAM
        % ($\lambda=0.05$, $\text{SNR}_{\text{Train}}=16dB$)}
        
        \addplot[green, solid, line width=0.9pt, 
        mark=*, mark options={fill=green, scale=1.1}, error bars/.cd, y dir=both, y explicit, every nth mark=1] 
                table [x=snr, y=nonuniform_klloss, 
                y error=nonuniform_klloss_std, col sep=comma]
                {data/djsccq_64qam_c256_tsnr16_psnr.csv};
        \addlegendentry{\textit{DeepJSCC-Q} 64-QAM 
        ($\lambda=0.05$, $\text{SNR}_{\text{Train}}=16dB$)}
        
        \addplot[orange, solid, line width=0.9pt, 
        mark=*, mark options={fill=orange, scale=1.1}, 
        error bars/.cd, y dir=both, y explicit, every nth mark=2] 
                table [x=snr, y=nonuniform,
                y error=nonuniform_std, col sep=comma]
                {data/djsccq_16qam_c256_tsnr16_psnr.csv};
        \addlegendentry{\textit{DeepJSCC-Q} 16-QAM
        ($\lambda=0.05$, $\text{SNR}_{\text{Train}}=16dB$)}

        \addplot[red, dashed, line width=0.9pt, 
        mark=*, mark options={fill=red, scale=1.1}, 
        error bars/.cd, y dir=both, y explicit, every nth mark=2] 
                table [x=snr, y=psnr, 
                y error=psnr_std, col sep=comma]
                {data/djscc_c256_tsnr16.csv};
        \addlegendentry{\textit{DeepJSCC} 
        ($\bar{\mathbf{z}}\in\mathbb{C}^k$, $||\bar{\mathbf{z}}||_2^2=k$)}
        \end{axis}
        \end{tikzpicture}
    }
    \hfill
  \subfloat[MS-SSIM\label{subfig:msssim_qam_v_noquant}]{%
    \begin{tikzpicture}
        \pgfplotsset{
            legend style={
                font=\fontsize{4}{4}\selectfont,
                at={(1.0,0.)},
                anchor=south east,
            },
            % height=0.4\linewidth,
            width=0.5\linewidth,
            xmin=0,
            xmax=17,
            ymin=0.86,
            ymax=1,
            xtick distance=2,
            ytick distance=0.02,
            xlabel={SNR (dB)},
            ylabel={MS-SSIM},
            grid=both,
            grid style={line width=.1pt, draw=gray!10},
            major grid style={line width=.2pt,draw=gray!50},
            every axis/.append style={
                x label style={
                    font=\fontsize{8}{8}\selectfont,
                    at={(axis description cs:0.5, -0.04)},
                    },
                y label style={
                    font=\fontsize{8}{8}\selectfont,
                    at={(axis description cs:-0.1,0.5)},
                    },
                x tick label style={
                    font=\fontsize{8}{8}\selectfont,
                    /pgf/number format/.cd,
                    fixed,
                    fixed zerofill,
                    precision=0,
                    /tikz/.cd
                    },
                y tick label style={
                    font=\fontsize{8}{8}\selectfont,
                    /pgf/number format/.cd,
                    fixed,
                    fixed zerofill,
                    precision=2,
                    /tikz/.cd
                    },
            }
        }
        \begin{axis}
        \addplot[blue, solid, line width=0.9pt, 
        mark=*, mark options={fill=blue, scale=1.1}, error bars/.cd, y dir=both, y explicit, every nth mark=1] 
                table [x=snr, y=soft_hard, 
                y error=soft_hard_std, col sep=comma]
                {data/djsccq_4096qam_c256_tsnr16_msssim.csv};
        \addlegendentry{\textit{DeepJSCC-Q} 4096-QAM 
        ($\lambda=0$, $\text{SNR}_{\text{Train}}=16dB$)}
        
        % \addplot[orange, dashed, line width=0.9pt, 
        % mark=*, mark options={fill=orange, scale=1.1}, error bars/.cd, y dir=both, y explicit, every nth mark=1] 
        %         table [x=snr, y=soft_hard_klloss,
        %         y error=nonuniform_klloss_std, col sep=comma]
        %         {data/djsccq_1024qam_c256_tsnr16_msssim.csv};
        % \addlegendentry{\textit{DeepJSCC-Q} 1024-QAM
        % ($\lambda=0.05$, $\text{SNR}_{\text{Train}}=16dB$)}
        
        \addplot[green, solid, line width=0.9pt, 
        mark=*, mark options={fill=green, scale=1.1}, error bars/.cd, y dir=both, y explicit, every nth mark=1] 
                table [x=snr, y=soft_hard, 
                y error=soft_hard_std, col sep=comma]
                {data/djsccq_64qam_c256_tsnr16_msssim.csv};
        \addlegendentry{\textit{DeepJSCC-Q} 64-QAM 
        ($\lambda=0.05$, $\text{SNR}_{\text{Train}}=16dB$)}
        
        \addplot[orange, solid, line width=0.9pt, 
        mark=*, mark options={fill=orange, scale=1.1}, error bars/.cd, y dir=both, y explicit, every nth mark=1] 
                table [x=snr, y=soft_hard,
                y error=soft_hard_std, col sep=comma]
                {data/djsccq_16qam_c256_tsnr16_msssim.csv};
        \addlegendentry{\textit{DeepJSCC-Q} 16-QAM
        ($\lambda=0.05$, $\text{SNR}_{\text{Train}}=16dB$)}
        
        \addplot[red, dashed, line width=0.9pt, 
        mark=*, mark options={fill=red, scale=1.1}, error bars/.cd, y dir=both, y explicit, every nth mark=1] 
                table [x=snr, y=msssim, 
                y error=msssim_std, col sep=comma]
                {data/djscc_c256_tsnr16.csv};
        \addlegendentry{\textit{DeepJSCC} 
        ($\bar{\mathbf{z}}\in\mathbb{C}^k$, $||\bar{\mathbf{z}}||_2^2=1$)}
        \end{axis}
        \end{tikzpicture}
        }
  \caption{
  Comparison of \emph{DeepJSCC-Q} with constellation orders $M \in \{16, 64, 4096\}$ against non-quantized \emph{DeepJSCC} for the AWGN channel case.
  }
  \label{fig:qam_v_noquant} 
\end{figure}

We note that the best performance with the separation approach at each channel SNR requires choosing the right constellation size.
Increasing the constellation order effectively increases the transmission rate.
However, while the increase in the transmission rate increases the quality of the compressed image, this will also increase the error probability over the channel.
Hence, we should choose the right constellation size for each channel SNR.
On the other hand, for \emph{DeepJSCC-Q}, given an SNR, the higher the constellation order, the better the end-to-end performance.
This can be understood from the perspective of quantization error.
A higher order modulation essentially corresponds to greater number of quantization levels of the encoder output $\mathbf{z} = f_{\boldsymbol{\theta}}(\mathbf{x})$, and thus a greater accuracy of $\bar{\mathbf{z}}$ in representing $\mathbf{z}$.
Since DeepJSCC has already been shown to surpass the performance of BPG and LDPC codes by \cite{Kurka:IZS2020}, naturally, as we increase the constellation order, the performance of \emph{DeepJSCC-Q} will approach that of DeepJSCC.
This is also why \emph{DeepJSCC-Q} with 4096-QAM constellation performs better than 64-QAM in Fig. \ref{fig:graceful} for $\text{SNR} > 7$, even when the 4096-QAM model was trained for a higher SNR than the 64-QAM model.

This observation is further supported by Fig. \ref{fig:qam_v_noquant}, where we can see that increasing the modulation order monotonically improves the performance of \emph{DeepJSCC-Q}.
Moreover, performance of \emph{DeepJSCC-Q} asymptotically approaches unconstrained DeepJSCC as the modulation order increases, with the $|\mathcal{C}|=4096$ model performing nearly the same.
Note that to compare to DeepJSCC, we simply remove the quantizer $q_\mathcal{C}$ and normalize the output power of the encoder, such that the channel input is
\begin{equation}
    \bar{\mathbf{z}} = \frac{\sqrt{kP}}{||\mathbf{z}||_2}\mathbf{z}.
\end{equation}
It is worth mentioning that for the modulation orders considered in Fig. \ref{fig:qam_v_noquant}, the LDPC codes considered herein cannot be decoded successfully for the SNR range tested.

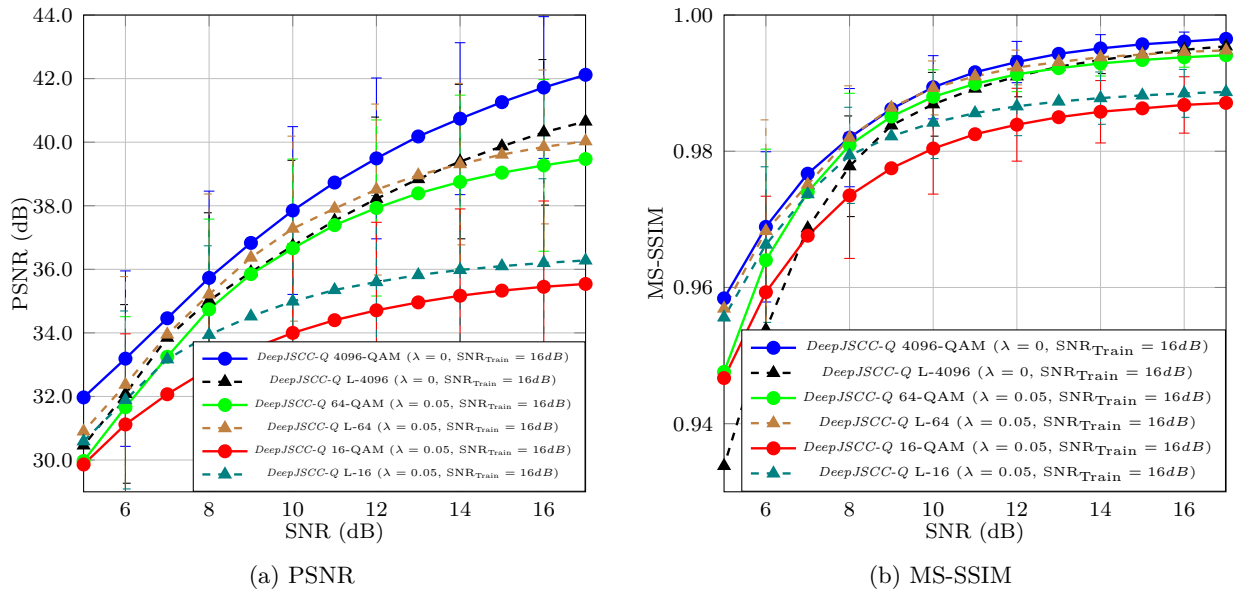
\begin{figure} 
    \centering
  \subfloat[PSNR\label{subfig:psnr_learned_v_qam}]{%
    \begin{tikzpicture}
        \pgfplotsset{
            legend style={
                font=\fontsize{4}{4}\selectfont,
                at={(1.0,0.0)},
                anchor=south east,
            },
            height=0.48\linewidth,
            width=0.5\linewidth,
            xmin=5,
            xmax=17,
            ymin=29,
            ymax=44,
            xtick distance=2,
            ytick distance=2,
            xlabel={SNR (dB)},
            ylabel={PSNR (dB)},
            grid=both,
            grid style={line width=.1pt, draw=gray!10},
            major grid style={line width=.2pt,draw=gray!50},
            every axis/.append style={
                x label style={
                    font=\fontsize{8}{8}\selectfont,
                    at={(axis description cs:0.5,-0.04)},
                    },
                y label style={
                    font=\fontsize{8}{8}\selectfont,
                    at={(axis description cs:-0.08,0.5)},
                    },
                x tick label style={
                    font=\fontsize{8}{8}\selectfont,
                    /pgf/number format/.cd,
                    fixed,
                    fixed zerofill,
                    precision=0,
                    /tikz/.cd
                    },
                y tick label style={
                    font=\fontsize{8}{8}\selectfont,
                    /pgf/number format/.cd,
                    fixed,
                    fixed zerofill,
                    precision=1,
                    /tikz/.cd
                    },
            }
        }
        \begin{axis}
        \addplot[blue, solid, line width=0.9pt, 
        mark=*, mark options={fill=blue, scale=1.1}, 
        error bars/.cd, y dir=both, y explicit, every nth mark=2] 
                table [x=snr, y=soft_hard, 
                y error=soft_hard_std, col sep=comma]
                {data/djsccq_4096qam_c256_tsnr16_psnr.csv};
        \addlegendentry{\textit{DeepJSCC-Q} 4096-QAM 
        ($\lambda=0$, $\text{SNR}_{\text{Train}}=16dB$)}
        
        \addplot[black, dashed, line width=0.9pt, 
        mark=triangle*, mark options={fill=black, scale=1.1, solid}, 
        error bars/.cd, y dir=both, y explicit, every nth mark=2] 
                table [x=snr, y=nonuniform, 
                y error=nonuniform_std, col sep=comma]
                {data/djsccq_4096qam_c256_tsnr16_psnr.csv};
        \addlegendentry{\textit{DeepJSCC-Q} L-4096 
        ($\lambda=0$, $\text{SNR}_{\text{Train}}=16dB$)}
        
        \addplot[green, solid, line width=0.9pt, 
        mark=*, mark options={fill=green, scale=1.1}, 
        error bars/.cd, y dir=both, y explicit, every nth mark=2] 
                table [x=snr, y=soft_hard_klloss, 
                y error=soft_hard_klloss_std, col sep=comma]
                {data/djsccq_64qam_c256_tsnr16_psnr.csv};
        \addlegendentry{\textit{DeepJSCC-Q} 64-QAM 
        ($\lambda=0.05$, $\text{SNR}_{\text{Train}}=16dB$)}
        
        \addplot[brown, dashed, line width=0.9pt, 
        mark=triangle*, mark options={fill=brown, scale=1.1, solid}, 
        error bars/.cd, y dir=both, y explicit, every nth mark=2] 
                table [x=snr, y=nonuniform_klloss, 
                y error=nonuniform_klloss_std, col sep=comma]
                {data/djsccq_64qam_c256_tsnr16_psnr.csv};
        \addlegendentry{\textit{DeepJSCC-Q} L-64
        ($\lambda=0.05$, $\text{SNR}_{\text{Train}}=16dB$)}
        
        \addplot[red, solid, line width=0.9pt, 
        mark=*, mark options={fill=red, scale=1.1}, 
        error bars/.cd, y dir=both, y explicit, every nth mark=2] 
                table [x=snr, y=soft_hard_klloss, 
                y error=soft_hard_klloss_std, col sep=comma]
                {data/djsccq_16qam_c256_tsnr16_psnr.csv};
        \addlegendentry{\textit{DeepJSCC-Q} 16-QAM 
        ($\lambda=0.05$, $\text{SNR}_{\text{Train}}=16dB$)}
        
        \addplot[teal, dashed, line width=0.9pt, 
        mark=triangle*, mark options={fill=teal, scale=1.1, solid}, 
        error bars/.cd, y dir=both, y explicit, every nth mark=2] 
                table [x=snr, y=nonuniform_klloss, 
                y error=nonuniform_klloss_std, col sep=comma]
                {data/djsccq_16qam_c256_tsnr16_psnr.csv};
        \addlegendentry{\textit{DeepJSCC-Q} L-16
        ($\lambda=0.05$, $\text{SNR}_{\text{Train}}=16dB$)}
        \end{axis}
        \end{tikzpicture}
    }
    \hfill
  \subfloat[MS-SSIM\label{subfig:msssim_learned_v_qam}]{%
    \begin{tikzpicture}
        \pgfplotsset{
            legend style={
                font=\fontsize{5}{5}\selectfont,
                at={(1.0,0.)},
                anchor=south east,
            },
            height=0.48\linewidth,
            width=0.5\linewidth,
            xmin=5,
            xmax=17,
            ymin=0.93,
            ymax=1,
            xtick distance=2,
            ytick distance=0.02,
            xlabel={SNR (dB)},
            ylabel={MS-SSIM},
            grid=both,
            grid style={line width=.1pt, draw=gray!10},
            major grid style={line width=.2pt,draw=gray!50},
            every axis/.append style={
                x label style={
                    font=\fontsize{8}{8}\selectfont,
                    at={(axis description cs:0.5, -0.04)},
                    },
                y label style={
                    font=\fontsize{8}{8}\selectfont,
                    at={(axis description cs:-0.1,0.5)},
                    },
                x tick label style={
                    font=\fontsize{8}{8}\selectfont,
                    /pgf/number format/.cd,
                    fixed,
                    fixed zerofill,
                    precision=0,
                    /tikz/.cd
                    },
                y tick label style={
                    font=\fontsize{8}{8}\selectfont,
                    /pgf/number format/.cd,
                    fixed,
                    fixed zerofill,
                    precision=2,
                    /tikz/.cd
                    },
            }
        }
        \begin{axis}%[mark options={solid}]
        \addplot[blue, solid, line width=0.9pt, 
        mark=*, mark options={fill=blue, scale=1.1}, 
        error bars/.cd, y dir=both, y explicit, every nth mark=2] 
                table [x=snr, y=soft_hard, 
                y error=soft_hard_std, col sep=comma]
                {data/djsccq_4096qam_c256_tsnr16_msssim.csv};
        \addlegendentry{\textit{DeepJSCC-Q} 4096-QAM 
        ($\lambda=0$, $\text{SNR}_{\text{Train}}=16dB$)}
        
        \addplot[black, dashed, line width=0.9pt, 
        mark=triangle*, mark options={fill=black, scale=1.1, solid}, 
        error bars/.cd, y dir=both, y explicit, every nth mark=2] 
                table [x=snr, y=nonuniform, 
                y error=nonuniform_std, col sep=comma]
                {data/djsccq_4096qam_c256_tsnr16_msssim.csv};
        \addlegendentry{\textit{DeepJSCC-Q} L-4096 
        ($\lambda=0$, $\text{SNR}_{\text{Train}}=16dB$)}
        
        \addplot[green, solid, line width=0.9pt, 
        mark=*, mark options={fill=green, scale=1.1}, 
        error bars/.cd, y dir=both, y explicit, every nth mark=2] 
                table [x=snr, y=soft_hard_klloss, 
                y error=soft_hard_klloss_std, col sep=comma]
                {data/djsccq_64qam_c256_tsnr16_msssim.csv};
        \addlegendentry{\textit{DeepJSCC-Q} 64-QAM 
        ($\lambda=0.05$, $\text{SNR}_{\text{Train}}=16dB$)}
        
        \addplot[brown, dashed, line width=0.9pt, 
        mark=triangle*, mark options={fill=brown, scale=1.1, solid}, 
        error bars/.cd, y dir=both, y explicit, every nth mark=2] 
                table [x=snr, y=nonuniform_klloss, 
                y error=nonuniform_std, col sep=comma]
                {data/djsccq_64qam_c256_tsnr16_msssim.csv};
        \addlegendentry{\textit{DeepJSCC-Q} L-64
        ($\lambda=0.05$, $\text{SNR}_{\text{Train}}=16dB$)}
        
        \addplot[red, solid, line width=0.9pt, 
        mark=*, mark options={fill=red, scale=1.1}, 
        error bars/.cd, y dir=both, y explicit, every nth mark=2] 
                table [x=snr, y=soft_hard_klloss, 
                y error=soft_hard_klloss_std, col sep=comma]
                {data/djsccq_16qam_c256_tsnr16_msssim.csv};
        \addlegendentry{\textit{DeepJSCC-Q} 16-QAM 
        ($\lambda=0.05$, $\text{SNR}_{\text{Train}}=16dB$)}
        
        \addplot[teal, dashed, line width=0.9pt, 
        mark=triangle*, mark options={fill=teal, scale=1.1, solid}, 
        error bars/.cd, y dir=both, y explicit, every nth mark=2] 
                table [x=snr, y=nonuniform_klloss, 
                y error=nonuniform_klloss_std, col sep=comma]
                {data/djsccq_16qam_c256_tsnr16_msssim.csv};
        \addlegendentry{\textit{DeepJSCC-Q} L-16
        ($\lambda=0.05$, $\text{SNR}_{\text{Train}}=16dB$)}
        \end{axis}
        \end{tikzpicture}
        }
  \caption{
  Comparison of \emph{DeepJSCC-Q} trained using fixed and learned constellation for the AWGN channel case.
  }
  \label{fig:learned_v_qam} 
\end{figure}

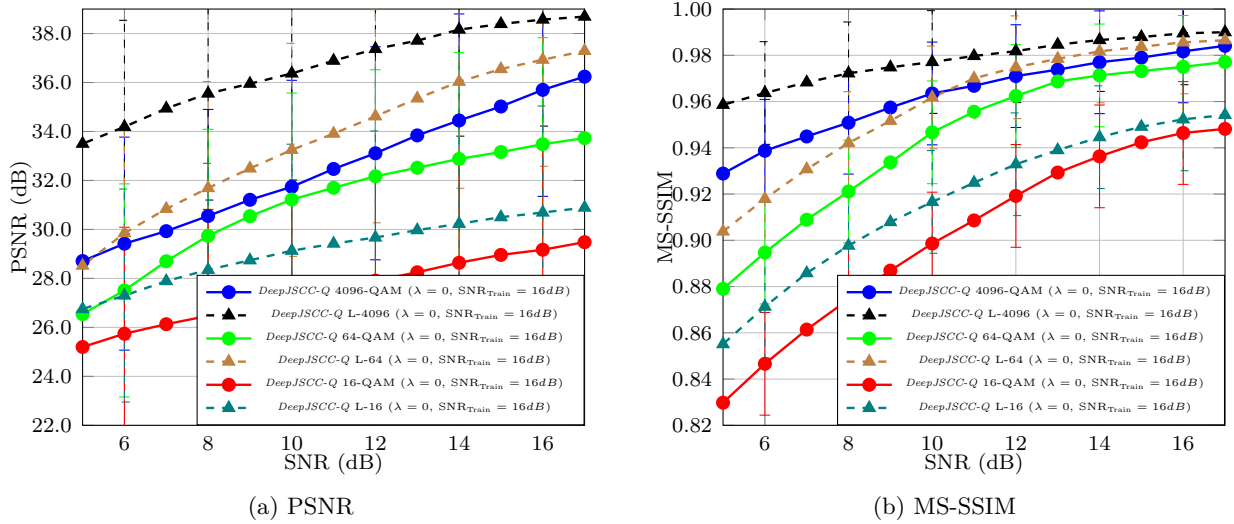
\begin{figure} 
    \centering
  \subfloat[PSNR\label{subfig:psnr_learned_v_qam_fading}]{%
    \begin{tikzpicture}
        \pgfplotsset{
            legend style={
                font=\fontsize{4}{4}\selectfont,
                at={(1.0,.0)},
                anchor=south east,
            },
            width=0.5\linewidth,
            xmin=5,
            xmax=17,
            ymin=22,
            ymax=39,
            xtick distance=2,
            ytick distance=2,
            xlabel={SNR (dB)},
            ylabel={PSNR (dB)},
            grid=both,
            grid style={line width=.1pt, draw=gray!10},
            major grid style={line width=.2pt,draw=gray!50},
            every axis/.append style={
                x label style={
                    font=\fontsize{8}{8}\selectfont,
                    at={(axis description cs:0.5,-0.04)},
                    },
                y label style={
                    font=\fontsize{8}{8}\selectfont,
                    at={(axis description cs:-0.08,0.5)},
                    },
                x tick label style={
                    font=\fontsize{8}{8}\selectfont,
                    /pgf/number format/.cd,
                    fixed,
                    fixed zerofill,
                    precision=0,
                    /tikz/.cd
                    },
                y tick label style={
                    font=\fontsize{8}{8}\selectfont,
                    /pgf/number format/.cd,
                    fixed,
                    fixed zerofill,
                    precision=1,
                    /tikz/.cd
                    },
            }
        }
        \begin{axis}
        \addplot[blue, solid, line width=0.9pt, 
        mark=*, mark options={fill=blue, scale=1.1}, 
        error bars/.cd, y dir=both, y explicit, every nth mark=2] 
                table [x=snr, y=soft_hard, 
                y error=soft_hard_std, col sep=comma]
                {data/djsccq_4096qam_c256_tsnr16_psnr_fading.csv};
        \addlegendentry{\textit{DeepJSCC-Q} 4096-QAM 
        ($\lambda=0$, $\text{SNR}_{\text{Train}}=16dB$)}
        
        \addplot[black, dashed, line width=0.9pt, 
        mark=triangle*, mark options={fill=black, scale=1.1, solid}, 
        error bars/.cd, y dir=both, y explicit, every nth mark=2] 
                table [x=snr, y=nonuniform, 
                y error=nonuniform_std, col sep=comma]
                {data/djsccq_4096qam_c256_tsnr16_psnr_fading.csv};
        \addlegendentry{\textit{DeepJSCC-Q} L-4096 
        ($\lambda=0$, $\text{SNR}_{\text{Train}}=16dB$)}
        
        \addplot[green, solid, line width=0.9pt, 
        mark=*, mark options={fill=green, scale=1.1}, 
        error bars/.cd, y dir=both, y explicit, every nth mark=2] 
                table [x=snr, y=soft_hard, 
                y error=soft_hard_std, col sep=comma]
                {data/djsccq_64qam_c256_tsnr16_psnr_fading.csv};
        \addlegendentry{\textit{DeepJSCC-Q} 64-QAM 
        ($\lambda=0$, $\text{SNR}_{\text{Train}}=16dB$)}
        
        \addplot[brown, dashed, line width=0.9pt, 
        mark=triangle*, mark options={fill=brown, scale=1.1, solid}, 
        error bars/.cd, y dir=both, y explicit, every nth mark=2] 
                table [x=snr, y=nonuniform, 
                y error=nonuniform_std, col sep=comma]
                {data/djsccq_64qam_c256_tsnr16_psnr_fading.csv};
        \addlegendentry{\textit{DeepJSCC-Q} L-64
        ($\lambda=0$, $\text{SNR}_{\text{Train}}=16dB$)}
        
        \addplot[red, solid, line width=0.9pt, 
        mark=*, mark options={fill=red, scale=1.1}, 
        error bars/.cd, y dir=both, y explicit, every nth mark=2] 
                table [x=snr, y=soft_hard, 
                y error=soft_hard_std, col sep=comma]
                {data/djsccq_16qam_c256_tsnr16_psnr_fading.csv};
        \addlegendentry{\textit{DeepJSCC-Q} 16-QAM 
        ($\lambda=0$, $\text{SNR}_{\text{Train}}=16dB$)}
        
        \addplot[teal, dashed, line width=0.9pt, 
        mark=triangle*, mark options={fill=teal, scale=1.1, solid}, 
        error bars/.cd, y dir=both, y explicit, every nth mark=2] 
                table [x=snr, y=nonuniform, 
                y error=nonuniform_std, col sep=comma]
                {data/djsccq_16qam_c256_tsnr16_psnr_fading.csv};
        \addlegendentry{\textit{DeepJSCC-Q} L-16
        ($\lambda=0$, $\text{SNR}_{\text{Train}}=16dB$)}
        \end{axis}
        \end{tikzpicture}
    }
    \hfill
  \subfloat[MS-SSIM\label{subfig:msssim_learned_v_qam_fading}]{%
    \begin{tikzpicture}
        \pgfplotsset{
            legend style={
                font=\fontsize{4}{4}\selectfont,
                at={(1.0,0.)},
                anchor=south east,
            },
            % height=0.4\linewidth,
            width=0.5\linewidth,
            xmin=5,
            xmax=17,
            ymin=0.82,
            ymax=1,
            xtick distance=2,
            ytick distance=0.02,
            xlabel={SNR (dB)},
            ylabel={MS-SSIM},
            grid=both,
            grid style={line width=.1pt, draw=gray!10},
            major grid style={line width=.2pt,draw=gray!50},
            every axis/.append style={
                x label style={
                    font=\fontsize{8}{8}\selectfont,
                    at={(axis description cs:0.5, -0.04)},
                    },
                y label style={
                    font=\fontsize{8}{8}\selectfont,
                    at={(axis description cs:-0.08,0.5)},
                    },
                x tick label style={
                    font=\fontsize{8}{8}\selectfont,
                    /pgf/number format/.cd,
                    fixed,
                    fixed zerofill,
                    precision=0,
                    /tikz/.cd
                    },
                y tick label style={
                    font=\fontsize{8}{8}\selectfont,
                    /pgf/number format/.cd,
                    fixed,
                    fixed zerofill,
                    precision=2,
                    /tikz/.cd
                    },
            }
        }
        \begin{axis}%[mark options={solid}]
        \addplot[blue, solid, line width=0.9pt, 
        mark=*, mark options={fill=blue, scale=1.1}, 
        error bars/.cd, y dir=both, y explicit, every nth mark=2] 
                table [x=snr, y=soft_hard, 
                y error=soft_hard_std, col sep=comma]
                {data/djsccq_4096qam_c256_tsnr16_msssim_fading.csv};
        \addlegendentry{\textit{DeepJSCC-Q} 4096-QAM 
        ($\lambda=0$, $\text{SNR}_{\text{Train}}=16dB$)}
        
        \addplot[black, dashed, line width=0.9pt, 
        mark=triangle*, mark options={fill=black, scale=1.1, solid}, 
        error bars/.cd, y dir=both, y explicit, every nth mark=2] 
                table [x=snr, y=nonuniform, 
                y error=nonuniform_std, col sep=comma]
                {data/djsccq_4096qam_c256_tsnr16_msssim_fading.csv};
        \addlegendentry{\textit{DeepJSCC-Q} L-4096 
        ($\lambda=0$, $\text{SNR}_{\text{Train}}=16dB$)}
        
        \addplot[green, solid, line width=0.9pt, 
        mark=*, mark options={fill=green, scale=1.1}, 
        error bars/.cd, y dir=both, y explicit, every nth mark=2] 
                table [x=snr, y=soft_hard, 
                y error=soft_hard_std, col sep=comma]
                {data/djsccq_64qam_c256_tsnr16_msssim_fading.csv};
        \addlegendentry{\textit{DeepJSCC-Q} 64-QAM 
        ($\lambda=0$, $\text{SNR}_{\text{Train}}=16dB$)}
        
        \addplot[brown, dashed, line width=0.9pt, 
        mark=triangle*, mark options={fill=brown, scale=1.1, solid}, 
        error bars/.cd, y dir=both, y explicit, every nth mark=2] 
                table [x=snr, y=nonuniform, 
                y error=nonuniform_std, col sep=comma]
                {data/djsccq_64qam_c256_tsnr16_msssim_fading.csv};
        \addlegendentry{\textit{DeepJSCC-Q} L-64
        ($\lambda=0$, $\text{SNR}_{\text{Train}}=16dB$)}
        
        \addplot[red, solid, line width=0.9pt, 
        mark=*, mark options={fill=red, scale=1.1}, 
        error bars/.cd, y dir=both, y explicit, every nth mark=2] 
                table [x=snr, y=soft_hard, 
                y error=soft_hard_std, col sep=comma]
                {data/djsccq_16qam_c256_tsnr16_msssim_fading.csv};
        \addlegendentry{\textit{DeepJSCC-Q} 16-QAM 
        ($\lambda=0$, $\text{SNR}_{\text{Train}}=16dB$)}
        
        \addplot[teal, dashed, line width=0.9pt, 
        mark=triangle*, mark options={fill=teal, scale=1.1, solid}, 
        error bars/.cd, y dir=both, y explicit, every nth mark=2] 
                table [x=snr, y=nonuniform, 
                y error=nonuniform_std, col sep=comma]
                {data/djsccq_16qam_c256_tsnr16_msssim_fading.csv};
        \addlegendentry{\textit{DeepJSCC-Q} L-16
        ($\lambda=0$, $\text{SNR}_{\text{Train}}=16dB$)}
        \end{axis}
        \end{tikzpicture}
        }
  \caption{
  Comparison of \emph{DeepJSCC-Q} trained using fixed and learned constellation for the slow fading channel case.
  }
  \label{fig:learned_v_qam_fading} 
\end{figure}

Next, we compare the models trained using soft-to-hard quantization with models trained using learned soft-to-hard quantization in Fig. \ref{fig:learned_v_qam} for the AWGN channel case.
We see that across $M = \{16, 64, 4096\}$, only the L-4096 result performed worse than its 4096-QAM counterpart.
This indicates that the degree of flexibility provided by the trainable constellation points can produce more coherent mapping between the source and channel input than a fixed M-QAM constellation.
The reason that the performance of the L-4096 model falls short of the 4096-QAM model is likely because the order of the modulation for $M = 4096$ is sufficiently high that the 4096-QAM constellation is already near optimal, and the large size of the constellation makes it challenging to further improve its performance.
This is supported by Fig. \ref{fig:qam_v_noquant}, where 4096-QAM performs nearly as well as the non-quantized result.
Although the L-4096 result should perform at least as good as the 4096-QAM result, it is possible that the objective function leads to large changes to the constellation that produces suboptimal results.
We believe that better tuning of the hyperparameters, such as the quantization hardness parameter $\sigma_q$, may alleviate this issue.
A similar pattern can be seen for the slow fading channel case shown in Fig. \ref{fig:learned_v_qam_fading}, where the margins between the learned and QAM constellations are much bigger, with the learned constellations producing substantially better results than their QAM counterparts.
In fact, the L-64 results even outperformed the 4096-QAM results in the PSNR metric.
This shows the importance of optimizing channel input geometry and distribution for non-Gaussian channels.

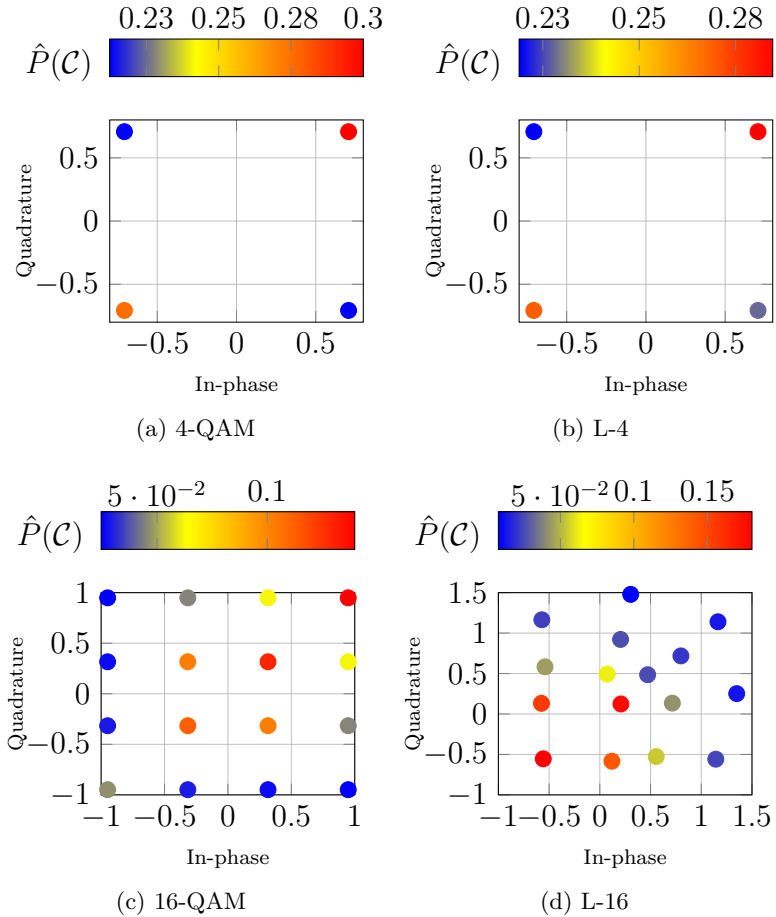
\begin{figure}
\centering
\subfloat[4-QAM \label{subfig:distribution_qpsk_soft_hard}]{
\begin{tikzpicture}
\pgfplotsset{
    % height=0.4\linewidth,
    width=0.3\textwidth,
    xmin=-0.8,
    xmax=0.8,
    ymin=-0.8,
    ymax=0.8,
    xtick distance=0.5,
    ytick distance=0.5,
    xlabel={In-phase},
    ylabel={Quadrature},
    grid=both,
    grid style={line width=.1pt, draw=gray!10},
    major grid style={line width=.2pt,draw=gray!50}
}
\begin{axis}[
    colorbar horizontal,
    colorbar style={
        at={(0,1.4)},
        anchor=north west,
        ylabel={$\hat{P}(\mathcal{C})$},
        ylabel style={anchor=north, rotate=270, at={(-0.2,1.2)}},
        xtick={0.2,0.225,0.25,0.275,0.3},
        xticklabel style={yshift=30pt}
    },
    y label style={
        font=\fontsize{8}{8}\selectfont,
        at={(axis description cs:-0.25,0.5)},
        },
    x label style={
        font=\fontsize{8}{8}\selectfont,
        },
]
\addplot[
  scatter,
  only marks,
  scatter src=explicit,
  mark=*,
  scatter/use mapped color={
    draw=mapped color,
    fill=mapped color,
  },
  scatter/@pre marker code/.append style=
    {/tikz/mark size={3pt}},
%   scatter/@pre marker code/.append style=
%     {/tikz/mark size={5pt+\pgfplotspointmetatransformed/650}},
    ]
  table[meta expr=abs(\thisrow{prob}), col sep=comma]
  {data/deepjscc_q_mod4_Qsoft_hard_klloss_bw0.1667_tsnr16_constellation.csv};
\end{axis}
\end{tikzpicture}
}
\subfloat[L-4 \label{subfig:distribution_qpsk_nonuniform}]{
\begin{tikzpicture}
\pgfplotsset{
    % height=0.4\linewidth,
    width=0.3\textwidth,
    xmin=-0.8,
    xmax=0.8,
    ymin=-0.8,
    ymax=0.8,
    xtick distance=0.5,
    ytick distance=0.5,
    xlabel={In-phase},
    ylabel={Quadrature},
    grid=both,
    grid style={line width=.1pt, draw=gray!10},
    major grid style={line width=.2pt,draw=gray!50}
}
\begin{axis}[
    colorbar horizontal,
    colorbar style={
        at={(0,1.4)},
        anchor=north west,
        ylabel={$\hat{P}(\mathcal{C})$},
        ylabel style={anchor=north, rotate=270, at={(-0.2,1.2)}},
        xtick={0.2,0.225,0.25,0.275,0.3},
        xticklabel style={yshift=30pt}
    },
    y label style={
        font=\fontsize{8}{8}\selectfont,
        at={(axis description cs:-0.25,0.5)},
        },
    x label style={
        font=\fontsize{8}{8}\selectfont,
        },
]
\addplot[
  scatter,
  only marks,
  scatter src=explicit,
  mark=*,
  scatter/use mapped color={
    draw=mapped color,
    fill=mapped color,
  },
  scatter/@pre marker code/.append style=
    {/tikz/mark size={3pt}},
    ]
  table[meta expr=abs(\thisrow{prob}), col sep=comma]
  {data/deepjscc_q_mod4_Qnonuniform_klloss_bw0.1667_tsnr16_constellation.csv};
\end{axis}
\end{tikzpicture}
}
\\
\subfloat[16-QAM \label{subfig:distribution_16qam_soft_hard}]{
\begin{tikzpicture}
\pgfplotsset{
    % height=0.4\linewidth,
    width=0.3\textwidth,
    xmin=-1,
    xmax=1,
    ymin=-1,
    ymax=1,
    xtick distance=0.5,
    ytick distance=0.5,
    xlabel={In-phase},
    ylabel={Quadrature},
    grid=both,
    grid style={line width=.1pt, draw=gray!10},
    major grid style={line width=.2pt,draw=gray!50}
}
\begin{axis}[
    colorbar horizontal,
    colorbar style={
        at={(0,1.4)},
        anchor=north west,
        ylabel={$\hat{P}(\mathcal{C})$},
        ylabel style={anchor=north, rotate=270, at={(-0.2,1.2)}},
        xtick={0.01,0.05,0.1,0.14},
        xticklabel style={yshift=30pt}
    },
    y label style={
        font=\fontsize{8}{8}\selectfont,
        at={(axis description cs:-0.25,0.5)},
        },
    x label style={
        font=\fontsize{8}{8}\selectfont,
        },
]
\addplot[
  scatter,
  only marks,
  scatter src=explicit,
  mark=*,
  scatter/use mapped color={
    draw=mapped color,
    fill=mapped color,
  },
  scatter/@pre marker code/.append style=
    {/tikz/mark size={3pt}},
    ]
  table[meta expr=abs(\thisrow{prob}), col sep=comma]
  {data/deepjscc_q_mod16_Qsoft_hard_klloss_bw0.1667_tsnr16_constellation.csv};
\end{axis}
\end{tikzpicture}
}
\subfloat[L-16 \label{subfig:distribution_16qam_nonuniform}]{
\begin{tikzpicture}
\pgfplotsset{
    % height=0.4\linewidth,
    width=0.3\textwidth,
    xmin=-1,
    xmax=1.5,
    ymin=-1,
    ymax=1.5,
    xtick distance=0.5,
    ytick distance=0.5,
    xlabel={In-phase},
    ylabel={Quadrature},
    grid=both,
    grid style={line width=.1pt, draw=gray!10},
    major grid style={line width=.2pt,draw=gray!50}
}
\begin{axis}[
    colorbar horizontal,
    colorbar style={
        at={(0,1.4)},
        anchor=north west,
        ylabel={$\hat{P}(\mathcal{C})$},
        ylabel style={anchor=north, rotate=270, at={(-0.2,1.2)}},
        xtick={0.05,0.1,0.15,0.2},
        xticklabel style={yshift=30pt}
    },
    y label style={
        font=\fontsize{8}{8}\selectfont,
        at={(axis description cs:-0.25,0.5)},
        },
    x label style={
        font=\fontsize{8}{8}\selectfont,
        },
]
\addplot[
  scatter,
  only marks,
  scatter src=explicit,
  mark=*,
  scatter/use mapped color={
    draw=mapped color,
    fill=mapped color,
  },
  scatter/@pre marker code/.append style=
    {/tikz/mark size={3pt}},
    ]
  table[meta expr=abs(\thisrow{prob}), col sep=comma]
  {data/deepjscc_q_mod16_Qnonuniform_klloss_bw0.1667_tsnr16_constellation.csv};
\end{axis}
\end{tikzpicture}
}
\caption{Probability distribution of constellation points for fixed and learned constellations in the AWGN channel case.}
\label{fig:constellation_distribution}
\end{figure}

To understand the type of constellations learned by \emph{DeepJSCC-Q} using learned soft-to-hard quantization, we refer to Fig. \ref{fig:constellation_distribution}, where we show the visualization of the constellation points and the probability of selecting each of the points for the AWGN channel case.
We can see that for a low modulation order, such as $M = 4$ shown in Fig. \ref{subfig:distribution_qpsk_soft_hard} and \ref{subfig:distribution_qpsk_nonuniform}, the constellation points learned by \emph{DeepJSCC-Q} are not very different from 4-QAM, and the distribution across the points is close to uniform.
However, when we increase the modulation order to $M = 16$, we begin to see significant differences between L-16 and 16-QAM, as shown by Fig. \ref{subfig:distribution_16qam_soft_hard} and \ref{subfig:distribution_16qam_nonuniform}.
The L-16 constellation is clearly non-square and not centered about the origin $(0, 0)$, with constellation points in the first quadrant having much higher power than those in the third quadrant.
This asymmetry may be part of the reason for the performance gain as the channel noise is zero mean and symmetric, meaning using the high power constellation points in the first quadrant can increase the instantaneous SNR of the received signal.
The average power is then maintained by choosing the constellations in the third quadrant much more frequently than the first quadrant.
Note that this observation can be seen as a form of unequal error protection (UEP) as well, with few important features using high power constellation points (first quadrant) and less important features using lower power constellation points (third quadrant).

\begin{figure}
\centering
\subfloat[16-QAM \label{subfig:distribution_16qam_soft_hard_fading}]{
\begin{tikzpicture}
\pgfplotsset{
    % height=0.4\linewidth,
    width=0.3\textwidth,
    xmin=-1,
    xmax=1,
    ymin=-1,
    ymax=1,
    xtick distance=0.5,
    ytick distance=0.5,
    xlabel={In-phase},
    ylabel={Quadrature},
    grid=both,
    grid style={line width=.1pt, draw=gray!10},
    major grid style={line width=.2pt,draw=gray!50}
}
\begin{axis}[
    colorbar horizontal,
    colorbar style={
        at={(0,1.4)},
        anchor=north west,
        ylabel={$\hat{P}(\mathcal{C})$},
        ylabel style={anchor=north, rotate=270, at={(-0.2,1.2)}},
        xtick={0.01,0.1,0.2},
        xticklabel style={yshift=30pt}
    },
    y label style={
        font=\fontsize{8}{8}\selectfont,
        at={(axis description cs:-0.25,0.5)},
        },
    x label style={
        font=\fontsize{8}{8}\selectfont,
        },
]
\addplot[
  scatter,
  only marks,
  scatter src=explicit,
  mark=*,
  scatter/use mapped color={
    draw=mapped color,
    fill=mapped color,
  },
  scatter/@pre marker code/.append style=
    {/tikz/mark size={3pt}},
    ]
  table[meta expr=abs(\thisrow{prob}), col sep=comma]
  {data/deepjscc_q_imagenet_mod16_Qsoft_hard_bw0.1667_fading_tsnr16_constellation.csv};
\end{axis}
\end{tikzpicture}
}
\subfloat[L-16 \label{subfig:distribution_16qam_nonuniform_fading}]{
\begin{tikzpicture}
\pgfplotsset{
    % height=0.4\linewidth,
    width=0.3\textwidth,
    xmin=-1,
    xmax=1.5,
    ymin=-1,
    ymax=1.5,
    xtick distance=0.5,
    ytick distance=0.5,
    xlabel={In-phase},
    ylabel={Quadrature},
    grid=both,
    grid style={line width=.1pt, draw=gray!10},
    major grid style={line width=.2pt,draw=gray!50}
}
\begin{axis}[
    colorbar horizontal,
    colorbar style={
        at={(0,1.4)},
        anchor=north west,
        ylabel={$\hat{P}(\mathcal{C})$},
        ylabel style={anchor=north, rotate=270, at={(-0.2,1.2)}},
        xtick={0.05,0.15,0.25},
        xticklabel style={yshift=30pt}
    },
    y label style={
        font=\fontsize{8}{8}\selectfont,
        at={(axis description cs:-0.25,0.5)},
        },
    x label style={
        font=\fontsize{8}{8}\selectfont,
        },
]
\addplot[
  scatter,
  only marks,
  scatter src=explicit,
  mark=*,
  scatter/use mapped color={
    draw=mapped color,
    fill=mapped color,
  },
  scatter/@pre marker code/.append style=
    {/tikz/mark size={3pt}},
    ]
  table[meta expr=abs(\thisrow{prob}), col sep=comma]
  {data/deepjscc_q_imagenet_mod16_Qnonuniform_bw0.1667_fading_tsnr16_constellation.csv};
\end{axis}
\end{tikzpicture}
}
\caption{Probability distribution of constellation points for fixed and learned constellations using modulation order $M=16$ in the slow fading channel case.}
\label{fig:constellation_distribution_fading}
\end{figure}
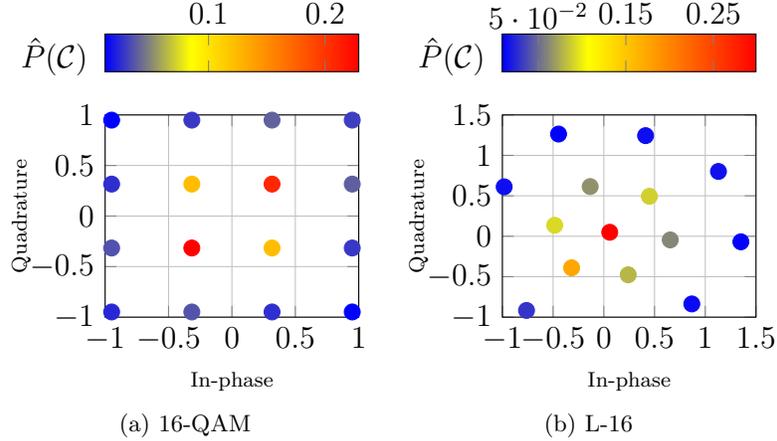

For the slow fading case,  we also see substantial differences between L-16 and 16-QAM, as shown in Fig. \ref{fig:constellation_distribution_fading}. 
While both the L-16 and the 16-QAM constellations select the central constellation points more frequently in the slow fading channel, the L-16 constellation is much more circular.
The L-16 constellation exhibits two circles around a center point.
The outer constellation points also have more power than even the highest power constellation points in 16-QAM, with the average power maintained by radially decreasing the probability distribution.
Moreover, when compared with the L-16 constellation learned under the AWGN channel, Fig. \ref{subfig:distribution_16qam_nonuniform_fading} shows a much more centered constellation, when compared to Fig. \ref{subfig:distribution_16qam_nonuniform}.
This is similar to the results from \cite{stark_joint_2019}, but here we are showing these properties in a JSCC context.

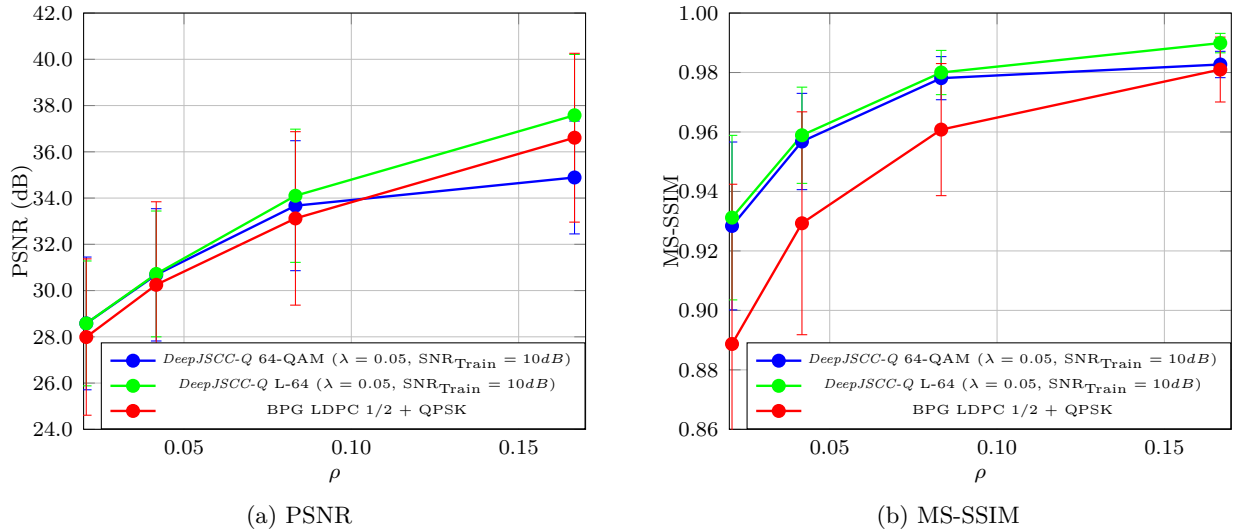
\begin{figure}
    \centering
  \subfloat[PSNR\label{subfig:psnr_v_bw}]{%
    \begin{tikzpicture}
        \pgfplotsset{
            legend style={
                font=\fontsize{5}{5}\selectfont,
                at={(1.0,.0)},
                anchor=south east,
            },
            width=0.5\linewidth,
            xmin=0.02,
            xmax=0.17,
            ymin=24,
            ymax=42,
            xtick distance=0.05,
            ytick distance=2,
            xlabel={$\rho$},
            ylabel={PSNR (dB)},
            grid=both,
            grid style={line width=.1pt, draw=gray!10},
            major grid style={line width=.2pt,draw=gray!50},
            every axis/.append style={
                x label style={
                    font=\fontsize{8}{8}\selectfont,
                    at={(axis description cs:0.5,-0.07)},
                    },
                y label style={
                    font=\fontsize{8}{8}\selectfont,
                    at={(axis description cs:-0.08,0.5)},
                    },
                x tick label style={
                    font=\fontsize{8}{8}\selectfont,
                    /pgf/number format/.cd,
                    fixed,
                    fixed zerofill,
                    precision=2,
                    /tikz/.cd
                    },
                y tick label style={
                    font=\fontsize{8}{8}\selectfont,
                    /pgf/number format/.cd,
                    fixed,
                    fixed zerofill,
                    precision=1,
                    /tikz/.cd
                    },
            }
        }
        \begin{axis}
        \addplot[blue, solid, line width=0.9pt, 
        mark=*, mark options={fill=blue, scale=1.1}, error bars/.cd, y dir=both, y explicit, every nth mark=1] 
                table [x=rho, y=soft_hard_klloss, 
                y error=soft_hard_klloss_std, col sep=comma]
                {data/rho_bw_psnr.csv};
        \addlegendentry{\textit{DeepJSCC-Q} 64-QAM
        ($\lambda=0.05$, $\text{SNR}_{\text{Train}}=10dB$)}
        
        \addplot[green, solid, line width=0.9pt, 
        mark=*, mark options={fill=green, scale=1.1}, error bars/.cd, y dir=both, y explicit, every nth mark=1] 
                table [x=rho, y=nonuniform_klloss, 
                y error=nonuniform_klloss_std, col sep=comma]
                {data/rho_bw_psnr.csv};
        \addlegendentry{\textit{DeepJSCC-Q} L-64
        ($\lambda=0.05$, $\text{SNR}_{\text{Train}}=10dB$)}
        
        \addplot[red, solid, line width=0.9pt, 
        mark=*, mark options={fill=red, scale=1.1}, error bars/.cd, y dir=both, y explicit, every nth mark=1] 
                table [x=rho, y=bpg, 
                y error=bpg_std, col sep=comma]
                {data/rho_bw_psnr.csv};
        \addlegendentry{BPG LDPC 1/2 + QPSK}
        \end{axis}
        \end{tikzpicture}
    }
    \hfill
  \subfloat[MS-SSIM\label{subfig:msssim_v_bw}]{%
    \begin{tikzpicture}
        \pgfplotsset{
            legend style={
                font=\fontsize{5}{5}\selectfont,
                at={(1.0,0.)},
                anchor=south east,
            },
            width=0.5\linewidth,
            xmin=0.02,
            xmax=0.17,
            ymin=0.86,
            ymax=1,
            xtick distance=0.05,
            ytick distance=0.02,
            xlabel={$\rho$},
            ylabel={MS-SSIM},
            grid=both,
            grid style={line width=.1pt, draw=gray!10},
            major grid style={line width=.2pt,draw=gray!50},
            every axis/.append style={
                x label style={
                    font=\fontsize{8}{8}\selectfont,
                    at={(axis description cs:0.5,-0.07)},
                    },
                y label style={
                    font=\fontsize{8}{8}\selectfont,
                    at={(axis description cs:-0.08,0.5)},
                    },
                x tick label style={
                    font=\fontsize{8}{8}\selectfont,
                    /pgf/number format/.cd,
                    fixed,
                    fixed zerofill,
                    precision=2,
                    /tikz/.cd
                    },
                y tick label style={
                    font=\fontsize{8}{8}\selectfont,
                    /pgf/number format/.cd,
                    fixed,
                    fixed zerofill,
                    precision=2,
                    /tikz/.cd
                    },
            }
        }
        \begin{axis}%[mark options={solid}]
        \addplot[blue, solid, line width=0.9pt, 
        mark=*, mark options={fill=blue, scale=1.1}, error bars/.cd, y dir=both, y explicit, every nth mark=1] 
                table [x=rho, y=soft_hard_klloss, 
                y error=soft_hard_klloss_std, col sep=comma]
                {data/rho_bw_msssim.csv};
        \addlegendentry{\textit{DeepJSCC-Q} 64-QAM
        ($\lambda=0.05$, $\text{SNR}_{\text{Train}}=10dB$)}
        
        \addplot[green, solid, line width=0.9pt, 
        mark=*, mark options={fill=green, scale=1.1}, error bars/.cd, y dir=both, y explicit, every nth mark=1] 
                table [x=rho, y=nonuniform_klloss, 
                y error=nonuniform_klloss_std, col sep=comma]
                {data/rho_bw_msssim.csv};
        \addlegendentry{\textit{DeepJSCC-Q} L-64
        ($\lambda=0.05$, $\text{SNR}_{\text{Train}}=10dB$)}
        
        \addplot[red, solid, line width=0.9pt, 
        mark=*, mark options={fill=red, scale=1.1}, error bars/.cd, y dir=both, y explicit, every nth mark=1] 
                table [x=rho, y=bpg, 
                y error=bpg_std, col sep=comma]
                {data/rho_bw_msssim.csv};
        \addlegendentry{BPG LDPC 1/2 + QPSK}
        \end{axis}
        \end{tikzpicture}
        }
  \caption{
  Comparison of \emph{DeepJSCC-Q} for different bandwidth compression ratio $\rho$ for the AWGN channel case.
  }
  \label{fig:distortion_v_bw} 
\end{figure}

\begin{figure}
\begin{center}
\includegraphics[width=\textwidth]{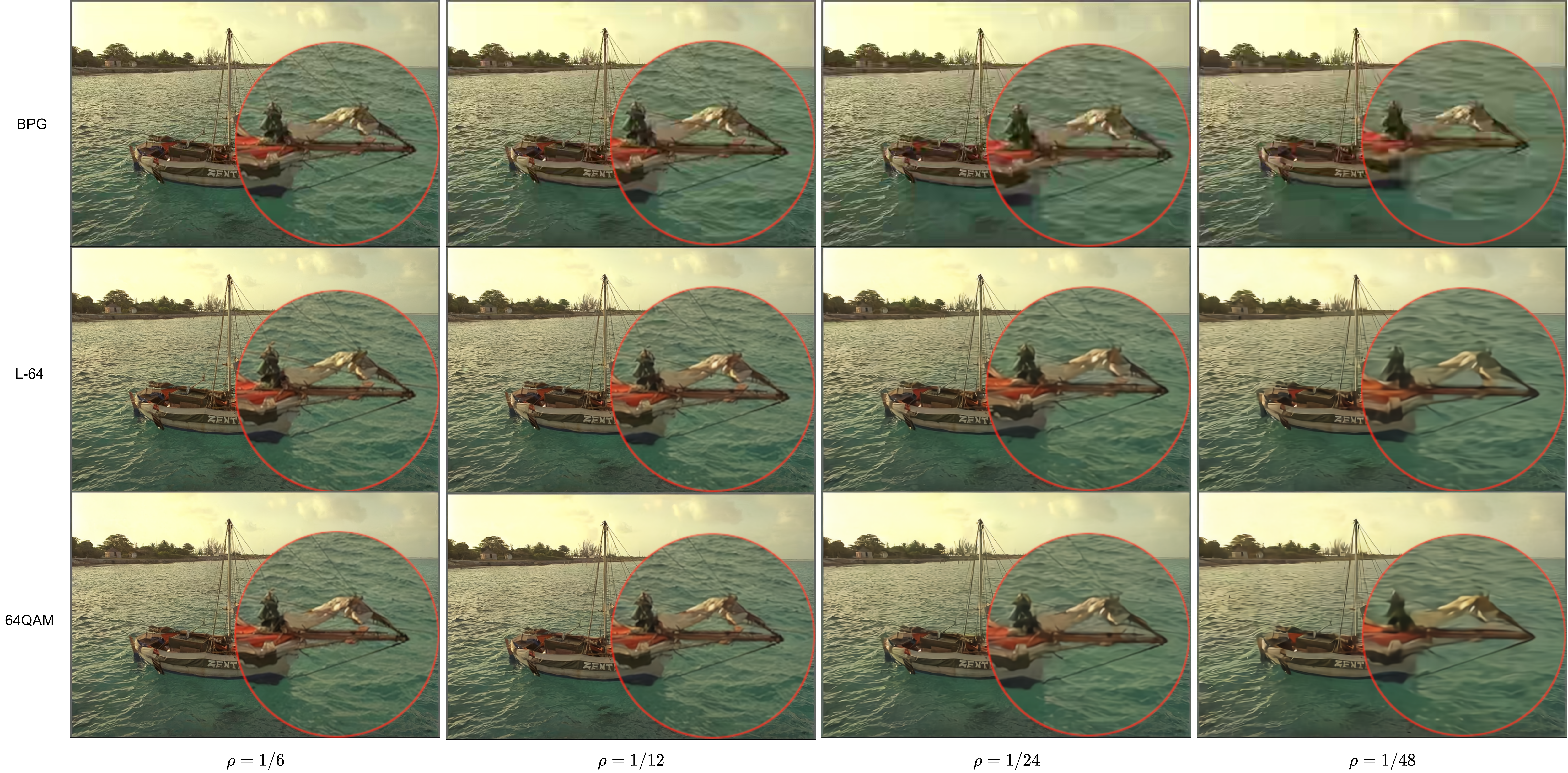}
\end{center}
  \caption{Visualization of image samples for different bandwidth compression ratios transmitted over the AWGN channel.}
\label{fig:visual_bw}
\end{figure}

Lastly, we investigate the relationship between the \emph{bandwidth compression ratio} $\rho$ and the reconstruction distortion.
Fig. \ref{fig:distortion_v_bw} compares the performance of \emph{DeepJSCC-Q} using $M = 64$ with separate source and channel coding for $\text{SNR} = 10$dB for the AWGN channel case.
This is akin to plotting the end-to-end rate distortion performance for each of the schemes considered herein.
We see that in all but one instance, \emph{DeepJSCC-Q} performs better than separation using BPG for source coding and an LDPC code for channel coding, demonstrating the superiority of JSCC in the finite block length regime.
Visual examples of the schemes considered herein given different bandwidth compression ratios $\rho$ are presented in Fig. \ref{fig:visual_bw}, where the last column of images clearly show that BPG produces more blurry images when compared to \emph{DeepJSCC-Q} using both L-64 and 64QAM constellations, under $\rho = 1/48$.
The only instance where \emph{DeepJSCC-Q} performed worse than separation is when the constellation is restricted to 64-QAM and the bandwidth compression ratio is $\rho = 1/6$.
This is likely because, as the bandwidth utilization increases, the encoder has greater degrees of freedom, allowing for more expressive channel input features that benefit from non-uniform quantization.

\section{Conclusions}
\label{sec:conclusions}

In this paper, we have proposed \emph{DeepJSCC-Q}, an end-to-end optimized joint source-channel coding scheme for wireless image transmission that is able to utilize a fixed channel input constellation and achieve similar performance to unquantized DeepJSCC, as previously proposed in \cite{Eirina:TCCN:19}.
Even with a constrained constellation, we are able to achieve superior performance to separation-based schemes using BPG for source coding and LDPC for channel coding, all the while avoiding the \textit{cliff-effect} that plagues the separation-based schemes.
This makes the viability of \emph{DeepJSCC-Q} in existing commercial hardware with standardized protocols much more attractive.
We also show that with sufficiently high modulation order, \emph{DeepJSCC-Q} can approach the performance of DeepJSCC, which does not have a fixed channel input constellation.
As such, if such constellations are available on the hardware, \emph{DeepJSCC-Q} can perform nearly as well as DeepJSCC.
Finally, we demonstrate that it is possible to learn a finite channel input alphabet that further improves the performance of \emph{DeepJSCC-Q}, with the resultant constellation showing highly non-trivial geometry and distribution.
These promising results can bring DNN-based JSCC schemes closer to real world deployment.

\bibliographystyle{ieeetr}
\bibliography{references.bib}

\end{document}